\documentclass[%
 reprint, prx,
 amsmath,amssymb,
 aps,
floatfix,
longbibliography
]{revtex4-2}

\usepackage{graphicx}% Include figure files
\usepackage{dcolumn}% Align table columns on decimal point
\usepackage{bm}% bold math
\begin{document}

\preprint{APS/123-QED}

\title{Spin excitations of a proximate Kitaev quantum spin liquid realized in Cu$_2$IrO$_3$}% Force line breaks with \\

\author{Sean K. Takahashi}
\author{Jiaming Wang}
\author{Alexandre Arsenault}

\affiliation{Department of Physics and Astronomy, McMaster University, Hamilton, Ontario, L8S 4M1, Canada}

%\affiliation{%


\author{Mykola Abramchuk}
\author{Fazel Tafti}
\affiliation{Department of Physics, Boston College, Chestnut Hill, Massachusetts 02467, United States}

\author{Philip M. Singer}
\affiliation{Department of Chemical and Biomolecular Engineering, Rice University, 6100 Main St., Houston, TX 77005-1892, United States}

\author{Takashi Imai} \email{imai@mcmaster.ca}
\affiliation{Department of Physics and Astronomy, McMaster University, Hamilton, Ontario, L8S 4M1, Canada}

%\affiliation{%


\date{\today}% It is always \today, today,

\begin{abstract}
Magnetic moments arranged at the corners of a honeycomb lattice are predicted to form a novel state of matter, Kitaev quantum spin liquid, under the influence of frustration effects between bond-dependent Ising interactions.  Some layered honeycomb iridates and related materials, such as Na$_{2}$IrO$_{3}$ and $\alpha$-RuCl$_{3}$, are proximate to Kitaev quantum spin liquid, but bosonic spin-wave excitations associated with undesirable antiferromagnetic long-range order mask the inherent properties of Kitaev Hamiltonian.  Here, we use $^{63}$Cu nuclear quadrupole resonance to uncover the low energy spin excitations in the nearly ideal honeycomb lattice of effective spin $S = 1/2$ at the Ir$^{4+}$ sites in Cu$_{2}$IrO$_{3}$.  We demonstrate that, unlike Na$_{2}$IrO$_{3}$, Ir spin fluctuations exhibit no evidence for critical slowing down toward magnetic long range order in zero external magnetic field.  Moreover, the low energy spin excitation spectrum is dominated by a mode that has a large excitation gap comparable to the Ising interactions, a signature expected for Majorana fermions of Kitaev quantum spin liquid.
\end{abstract}

%\begin{abstract}
%An article usually includes an abstract, a concise summary of the work
%covered at length in the main body of the article. 
%\begin{description}
%\end{description}
%\end{abstract}

\maketitle

%\tableofcontents

\section{Introduction}
Quantum spin liquids (QSL's) are a liquid-like state of matter, in which entangled spins remain paramagnetic without undergoing a magnetic long-range order.  Identifying a QSL is the holy grail of quantum condensed matter research today \cite{Lee2008, Balents2010, Rau2016, Imai2016}.  A recent milestone in the search of QSL's is the rigorous theoretical solution of Kitaev QSL for effective spin $S = 1/2$ arranged in a honeycomb lattice \cite{Kitaev2006}.  The Kitaev Hamiltonian may be written as
\begin{equation}
H_{K} = \sum_{<i, j>}J_{K}^{\gamma}S_{i}^{\gamma}S_{j}^{\gamma},
\end{equation} 
where $J_{K}^{\gamma}$ ($\gamma = x$, $y$, and $z$) represents the bond-dependent, frustrated Ising interaction as schematically shown in Fig. 1A, and the lattice summation $<i, j>$ is taken between nearest-neighbor spins.  Kitaev proved based on analytic calculations that his Hamiltonian supports a quantum spin liquid ground state. 

Besides possessing the ground state with no magnetic long range order, ideal Kitaev QSL is known to exhibit a variety of exotic properties \cite{Kitaev2006, Knolle2015, YoshitakePRL2016, Yoshitake2017, Yoshitake2}.  For example, fractionalization of quantum spins leads to two types of Majorana fermions, itinerant Majorana fermions and localized fluxes.  Therefore, elementary excitations in Kitaev QSL are not conventional magnons (spin waves), and may exhibit an inherent gap in its spin excitation spectrum, depending on the anisotropy of the Ising interactions in eq.(1).  Thermodynamic properties could show non-trivial behavior, too.  For example, the uniform spin susceptibility $\chi_{spin}$ is expected to increase or saturate with decreasing temperature, even if the excitation spectrum of Majorana fermions are inherently gapped \cite{YoshitakePRL2016, Yoshitake2017, Yoshitake2};  this is due to the non-conservative nature of the z-component of the total spin, $S_z$.

The Ir$^{4+}$, Rh$^{4+}$, and Ru$^{3+}$ ions in Na$_2$IrO$_3$ \cite{Singh2010}, Li$_2$RhO$_3$ \cite{Khunita2017}, 
$\alpha$-RuCl$_3$ \cite{Sandilands2015}, and more recently discovered H$_3$LiIr$_2$O$_6$\cite{Kitagawa2018} are under the influence of spin-orbit interactions, possess effective spin $S_{i} = 1/2$, and interact with each other primarily via frustrated Ising interactions \cite{Jackeli2009}.  Therefore, the honeycomb planes formed by these spins have been proposed to harbor Kitaev QSL \cite{Jackeli2009,Chaloupka2010,Chaloupka2013}.   In reality, additional interactions, such as Heisenberg's exchange term $H_{H} = \sum_{<i, j>}J_{H}{\bf S}_{i}\cdot\bf{S}_{j}$ and the symmetric off-diagonal exchange term $H_{\Gamma} =\sum_{<i, j>}\Gamma (S_{i}^{\alpha}S_{j}^{\beta} + S_{i}^{\beta}S_{j}^{\alpha})$ \cite{Rau2016,Jackeli2009,Chaloupka2010} perturb Kitaev Hamiltonian in eq.(1), leading to more complicated Kitaev-Heisenberg Hamiltonian,
\begin{equation}
H_{KH} = H_{K} + H_{H} + H_{\Gamma}.
\end{equation} 
Theoretical calculations established that Kitaev-Heisenberg Hamiltonian has a long-range ordered ground state rather than the spin liquid state in a broad parameter space \cite{Rau2016}.  Experimentally, Na$_2$IrO$_3$ \cite{Singh2010}, the first model material proposed for Kitaev-Heisenberg model \cite{Chaloupka2010}, indeed has a zig-zag antiferromagnetic long-range order below $T_{N} \sim 17$ K, as shown schematically in Fig. 1B \cite{Liu2011,Choi2012}.  Such a long-range ordered ground state conceals the inherent properties of Kitaev QSL.  In particular, bosonic spin-wave excitations (with total spin 1) associated with the undesirable antiferromagnetic long-range order dominate  the low energy sector of spin excitation spectrum  in Na$_2$IrO$_3$ \cite{Choi2012} and $\alpha$-RuCl$_3$ \cite{Banerjee2016,Banerjee2017Science,Banerjee2018}, as shown in the left panel of Figure 2A.  This makes it difficult to  identify the intrinsic excitations expected for eq.(1), {\it i.e.} the fractionalized Majorana fermions (with total spin 1/2) and pairs of gauge fluxes (with total spin 1).  Untangling the influence of $H_{H}$ and $H_{\Gamma}$ on $H_{K}$ has been at the focus of intense research both theoretically and experimentally, but there have been only a handful of proximate Kitaev materials known to date.

\begin{figure}
	\begin{center}
		\includegraphics[width=3.2in]{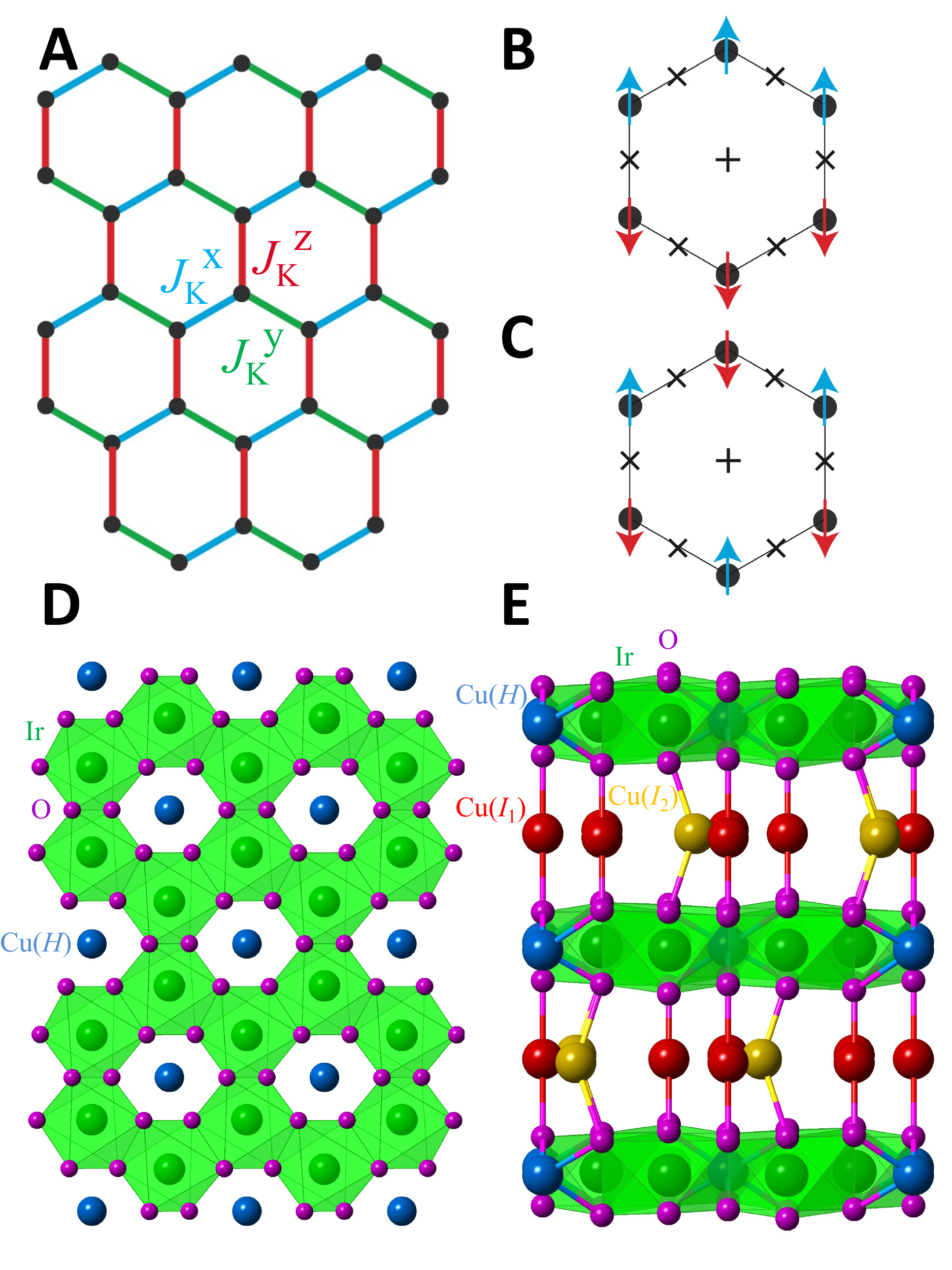}
		\caption{(A) Kitaev lattice of effective spin $S = 1/2$'s (represented by bullets) at the corners of a honeycomb structure.  Each spin interacts with three neighboring spins with Ising interactions $J_{K}^{x}$ (blue bonds), $J_{K}^{y}$ (green bonds), and $J_{K}^{z}$ (red bonds).  The frustration effects between the bonds lead to formation of Kitaev quantum spin liquid.  Possible long-range ordering patterns of Ir spins \cite{Liu2011,Choi2012} in (B) zig-zag and (C) N\'eel structure.  + symbols show the location of the Cu(H) sites, while $\times$ symbols indicate the approximate location of Cu(I$_{1}$) sites above or below the honeycomb planes.  (D) Honeycomb lattice formed by Ir$^{4+}$ ions in Cu$_2$IrO$_3$, viewed along the crystal c*-axis.  Each Ir$^{4+}$ (green) is surrounded by six O$^{2-}$ ions (purple) at the corners of an octahedron.  Cu$^{+}$(H) (light blue) is at the center of each iridium hexagon.  (E) Side view of the Ir$^{4+}$ planes viewed along the crystal a-axis, with the inter-planar Cu$^{+}$(I$_1$) (red) and Cu$^{+}$(I$_{2}$) (yellow) sites. O-Cu(I$_{1}$)-O bond is straight, whereas O-Cu(I$_{2}$)-O bond is buckled.}
		\label{fig:structure}
	\end{center}
\end{figure}

\begin{figure}
	\begin{center}
		\includegraphics[width=3.2in]{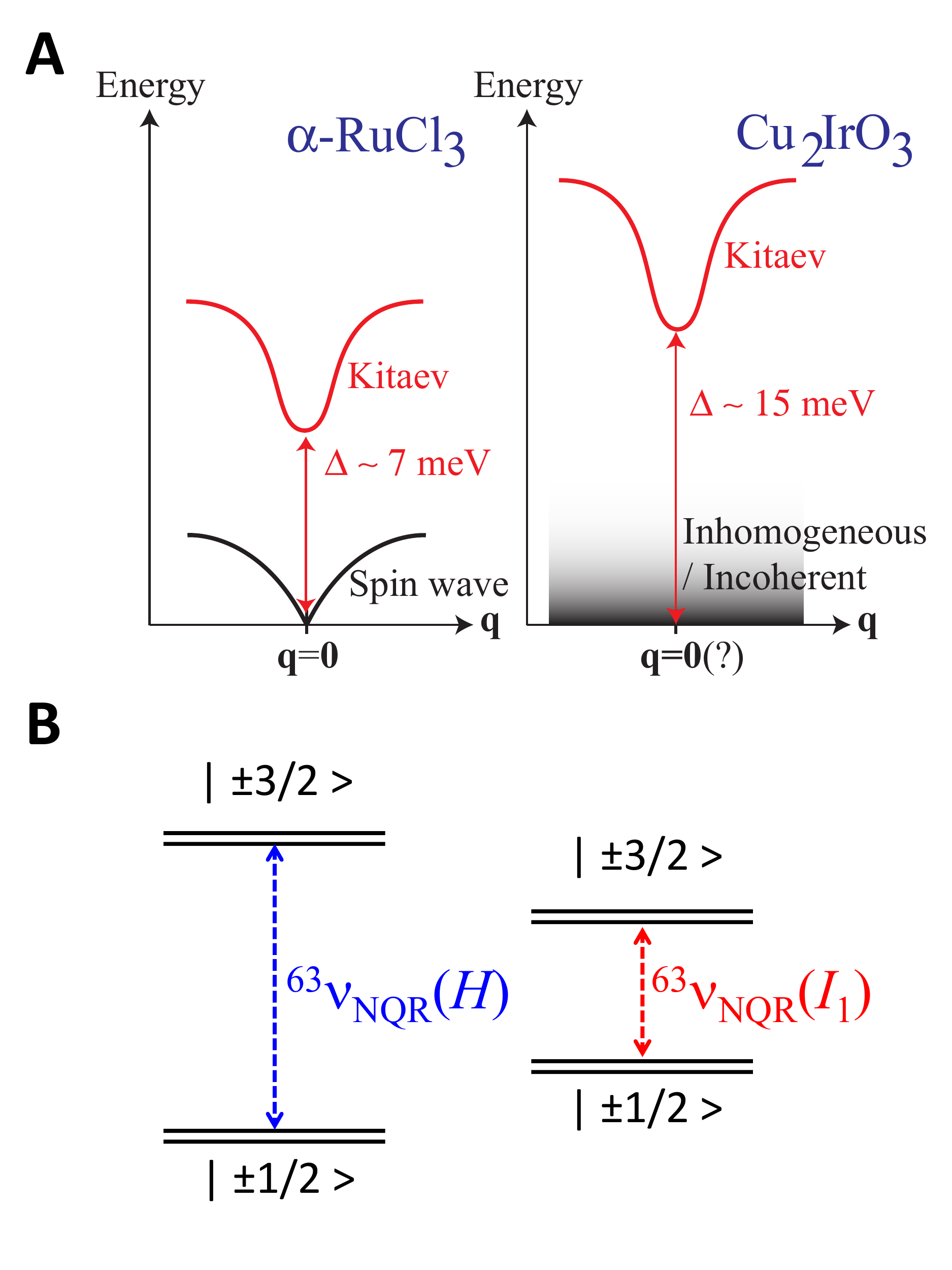}
		\caption{
		(A) Conceptual diagrams of the spin excitation spectrum in Kitaev candidate material $\alpha$-RuCl$_{3}$ with additional spin-wave like modes below the gap $\Delta \sim 7$~meV \cite{Banerjee2016, Banerjee2017Science,Banerjee2018} (left), and Cu$_2$IrO$_3$ with only spatially inhomogeneous, incoherent spin excitations at low energies (right).  (B) Schematic diagram of the quadrupolar splitting between the nuclear spin $|\pm1/2>$ and $|\pm3/2>$ states for $^{63}$Cu(H) and $^{63}$Cu(I$_1$) sites. 
		}
		\label{fig:excitations}
	\end{center}
\end{figure}
	
	Recent discovery of Cu$_2$IrO$_3$ \cite{Abramchuk2017} by substitution of Cu$^+$ ions into the Na$^+$ sites of Na$_2$IrO$_3$ paved a new avenue to investigate Kitaev QSL.  We present the crystal structure of Cu$_2$IrO$_3$ in Figure 1D-E \cite{Abramchuk2017}.  Both the Ir-Ir-Ir  (118.7$^\circ$ to 122.5$^\circ$) and Ir-O-Ir (95$^\circ$ to 98$^\circ$) bond angles are closer to the ideal Kitaev geometry of 120$^\circ$ and 90$^\circ$ \cite{Jackeli2009,Chaloupka2010}, respectively, than those of Na$_2$IrO$_3$ (114.9$^\circ$ to 124.2$^\circ$ and 98$^\circ$ to 99.4$^\circ$, respectively \cite{Singh2010}).  There are two distinct types of Cu sites in Cu$_2$IrO$_3$: Cu(H) sites are located within the honeycomb plane in the middle of each hexagon formed by six Ir sites, while inter-planar Cu(I) sites sit halfway between honeycomb planes (i.e. Cu(I)$_{1.5}$Cu(H)$_{0.5}$IrO$_3$).  The uniform magnetic susceptibility $\chi$ observed for Cu$_2$IrO$_3$ is nearly identical with that of Na$_2$IrO$_3$ except below $\sim$30 K, where defect spins dominate $\chi$ in Cu$_2$IrO$_3$ \cite{Abramchuk2017, Choi2018}.  In the case of Na$_2$IrO$_3$,	The theoretical fit of $\chi$ based on Kitaev-Heisenberg Hamiltonian was very good, and led to the initial estimate of $J_{K} \sim 21$~ meV and $J_{H} \sim -4$~meV \cite{Chaloupka2010}.  In order to account for the zig-zag ordered ground state, however, the additional off-diagonal exchange terms in eq.(2) are required \cite{Rau2016}.  Subsequent first principles calculations based on exact diagonalization, quantum chemistry method, and perturbation theory led to $J_{K} = -16.8 \sim -29.4$~ meV, $J_{H}  = 0.5\sim 3.2$~meV, and $\Gamma \sim 1$~meV \cite{Winter2017}.  In view of the nearly identical $\chi$, we expect that spin-spin interactions in Cu$_2$IrO$_3$ are comparable to these estimations.
	
	 Unlike Na$_2$IrO$_3$, however, Cu$_2$IrO$_3$ shows no evidence for antiferromagnetic long-range order \cite{Abramchuk2017}, and recent $\mu$SR \cite{Kenney2018,Choi2018} as well as our NQR measurements presented below establish that only $\sim$50\% of the sample volume exhibits spin freezing below the onset temperature of $T_{f} \sim10$~K.  Moreover, muons exhibit no spin precession about static hyperfine field, and Ir spins remain dynamic even below 1 K \cite{Kenney2018,Choi2018}.   This means that the low energy sector of the spin excitation spectrum in Cu$_2$IrO$_3$ is largely immune from the spin-waves, as schematically shown in the right panel of Figure 2A, and opens a wide temperature window above $T_f$ to probe the inherent properties  of Kitaev QSL. 
	 
On the other hand, Cu$_2$IrO$_3$ has its own complications arising from stacking faults, as is often the case for previously identified Kitaev candidate materials, and from the aforementioned defect spins  \cite{Kenney2018}.  The defect-induced enhanced susceptibility below $\sim 30$~K is known to obey a peculiar scaling law \cite{Choi2018}, reminiscent of analogous scaling behavior reported earlier for other spin liquid candidate materials with disorder \cite{Kimchi2018}.  The origin and nature of the defect spins in Cu$_2$IrO$_3$ as well as in  other related iridates \cite{Manni2014, Manni2014_2,Wallace2015, Kitagawa2018} are a matter of intense research and yet to be clarified.  In the case of Cu$_2$IrO$_3$, XANES measurements showed that up to $\sim$1/3 of $^{63}$Cu(H) sites are occupied by Cu$^{2+}$ ions rather than Cu$^+$ ions \cite{Kenney2018}.  To maintain charge neutrality, the charge state of an Ir$^{4+}$ ion(s) in their vicinity may change, possibly to Ir$^{3+}$ with null spin.  If randomly distributed, they might create bond disorder \cite{Knolle2019} and/or site dilution in the Kitaev Hamiltonian, in addition to the decoration of the Kitaev lattice with additional spin $S=1/2$ at the Cu$^{2+}$ sites.  Curiously, the bulk susceptibility of Cu$_2$IrO$_3$ and Na$_2$IrO$_3$ shares nearly identical Weiss temperature $\Theta_{w} =-110 \sim -123$~K \cite{Abramchuk2017}, whereas Mg$^{2+}$ substitution into Na$_2$IrO$_3$ is known to suppress $\Theta_{w}$ quickly to almost zero \cite{Wallace2015}.
	
	In this paper, we will use $^{63}$Cu nuclear quadrupole resonance (NQR), a variant of NMR, to probe intrinsic spin excitations in the paramagnetic state of Ir honeycomb layers in Cu$_2$IrO$_3$.  A major thrust of our work is that we can probe the Ir honeycomb lattice using NQR in zero magnetic field, owing to the strong nuclear quadrupole interaction at the Cu sites.  Moreover, the magnetic hyperfine coupling $A_{hf}$ between Cu nuclear spins and Ir electron spins via the transferred spin polarization into the empty Cu 4s orbital \cite{Mila-Rice} is generally strong.  Accordingly, defect spins do not overwhelm the intrinsic $^{63}$Cu NQR and NMR properties of Cu$_2$IrO$_3$, unlike $^1$H NMR data for the kagome lattice in herbertsmithite ZnCu$_3$(OH)$_6$Cl$_2$ \cite{Imai2008,Imai2012,Fu2015} and for the honeycomb lattice of H$_3$LiIr$_2$O$_6$ \cite{Kitagawa2018}.  We will demonstrate that Cu$_2$IrO$_3$ remains paramagnetic, and exhibits a large gap $\Delta$ comparable to the magnitude of Ising interaction, $\Delta \sim 15$ meV ($ = 0.5 | J_{K} | \sim 0.7 |J_{K}| $) in the primary channel of spin excitations {\it in zero external magnetic field}, $B_{ext}=$0.    
	
\section{Experimental Methods}

We synthesized the powder samples of Na$_2$IrO$_3$, Cu$_2$IrO$_3$ and Cu$_{1.5}$Na$_{0.5}$SnO$_{3}$ for NMR measurements via solid state reaction and ion exchange as described in \cite{Abramchuk2017, Abramchuk2018}.  We conducted all NMR and NQR measurements using home-built NMR spectrometers with the standard $\pi$/2 - $\pi$ spin echo pulse sequence. The typical pulse width for the $\pi$/2 pulse was 2 to 7 $\mu$s, and the pulse separation time $\tau \sim 10$ $\mu$s.  We measured $1/T_{1}$ using the inversion recovery method whereby we applied a $\pi$ pulse, followed by a spin echo sequence for various delay times $t$.  We fitted the inversion recovery of nuclear magnetization $M(t)$ to the exponential function expected for the observed transition \cite{Andrew1961, Narath1967}, as discussed in detail in Appendix A.  We measured the NQR intensity near and below $T_f$ at the peak of the Cu(I$_1$) and Cu(H) sites using a fixed delay time $\tau = 10$ $\mu$s.  We uni-axially aligned the powder sample of Cu$_2$IrO$_3$ in Stycast 1266 \cite{Takigawa1989} by curing the mixture for ~12 hours in a mold made of Teflon in a magnetic field of 9 T. 

\section{NMR Results}
\subsection{$^{63,65}$Cu NQR lineshapes}
	Naturally occurring $^{63,65}$Cu isotopes at Cu(H) and Cu(I) sites have nuclear spin $I = 3/2$ accompanied by a nuclear quadrupole moment $^{63,65}Q$, and the latter interacts with the electric field gradient (EFG) generated by electrons and ions in the lattice.  As shown schematically in Figs.\ 2B, this nuclear quadrupole interaction lifts the degeneracy of the nuclear spin $|\pm1/2>$ and $|\pm3/2>$ states even in the absence of an external magnetic field $B_{ext}$.  It is therefore possible to conduct $^{63,65}$Cu NMR based on the NQR techniques in $B_{ext}=0$ without perturbing the Ir electron spins with Zeeman effects, a crucial advantage of Cu$_2$IrO$_3$ for NMR investigation into the inherent Kitaev physics.  
	
	In Figs.\ 3A-B, we compare the $^{63,65}$Cu NQR lineshapes observed for Cu$_2$IrO$_3$ and a non-magnetic reference material Cu$_{1.5}$Na$_{0.5}$SnO$_{3}$ \cite{Abramchuk2018} with Na$_{0.5}$Sn honeycomb layers.  The NQR peak frequency $^{63,65}\nu_{NQR}$ of the $^{63,65}$Cu isotopes is proportional to $^{63,65}Q$, where $^{65}Q/^{63}Q =$ 0.927, and hence each $^{63}$Cu NQR peak observed at $^{63}\nu_{NQR}$ is accompanied by a smaller $^{65}$Cu NQR peak at a lower frequency $^{65}\nu_{NQR}=0.927 \cdot~^{63}\nu_{NQR}$ with the intensity ratio set by their natural abundance, 69\% and 31\%.  The two pairs of similar $^{63,65}$Cu NQR peaks observed below 30~MHz for both Cu$_2$IrO$_3$ and Cu$_{1.5}$Na$_{0.5}$SnO$_{3}$ indicate that they arise from two distinct types of the inter-layer Cu(I$_1$) and Cu(I$_2$) sites (the O-Cu(I$_2$)-O bond has a kink, see Figure 1E).  We assign the remaining $^{63,65}$Cu NQR peaks observed at around 52~MHz to the honeycomb Cu(H) sites.  

We note that the Kitaev candidate materials tend to suffer from stacking faults.  In the case of Cu$_2$IrO$_3$, due to the weak Cu-O-Cu bonds between the layers of Cu$_2$IrO$_3$, the material is prone to stacking faults in the form of twinning between the adjacent layers \cite{Abramchuk2018}.  The stacking faults are extended disorders that happen over several unit cells, however, we have to effectively model them in one unit cell when we solve the crystal structure by Rietveld refinements.  The result is two inequivalent dumbbell bonds, one straight and one twisted. If both bonds were perfectly straight, there would have been no twisting between adjacent layers and no stacking disorder.  Our observation of two distinct types of dumbbell Cu(I$_1$) and Cu(I$_2$) sites supports this.

\begin{figure}
	\begin{center}
		\includegraphics[width=3.2in]{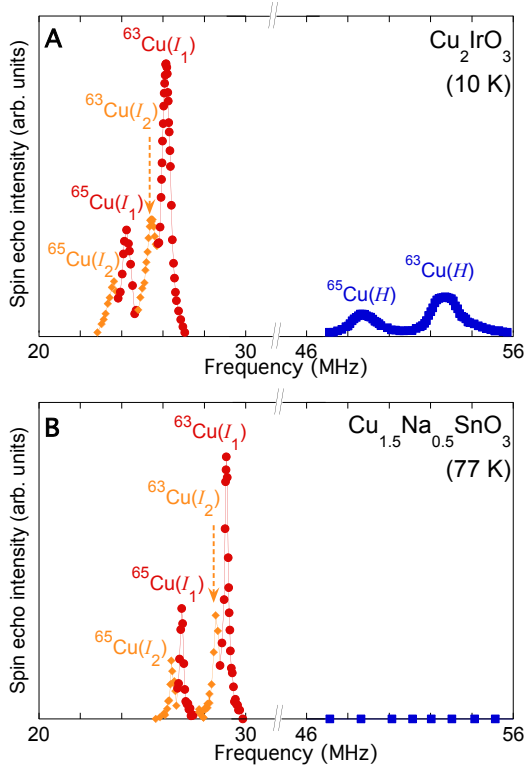}
		\caption{(A) $^{63,65}$Cu NQR lineshape observed at 10 K for the in-plane Cu(H) honeycomb site and the inter-planar Cu(I$_1$) and Cu(I$_2$) sites of Cu$_2$IrO$_3$.  (B) $^{63,65}$Cu NQR lineshape observed at 77 K for non-magnetic reference sample Cu$_{1.5}$Na$_{0.5}$SnO$_{3}$, which lacks Cu(H) sites.}
		\label{fig:NQRspectra}
	\end{center}
\end{figure}

\begin{figure}
	\begin{center}
		\hspace*{0.2in}\includegraphics[width=3in]{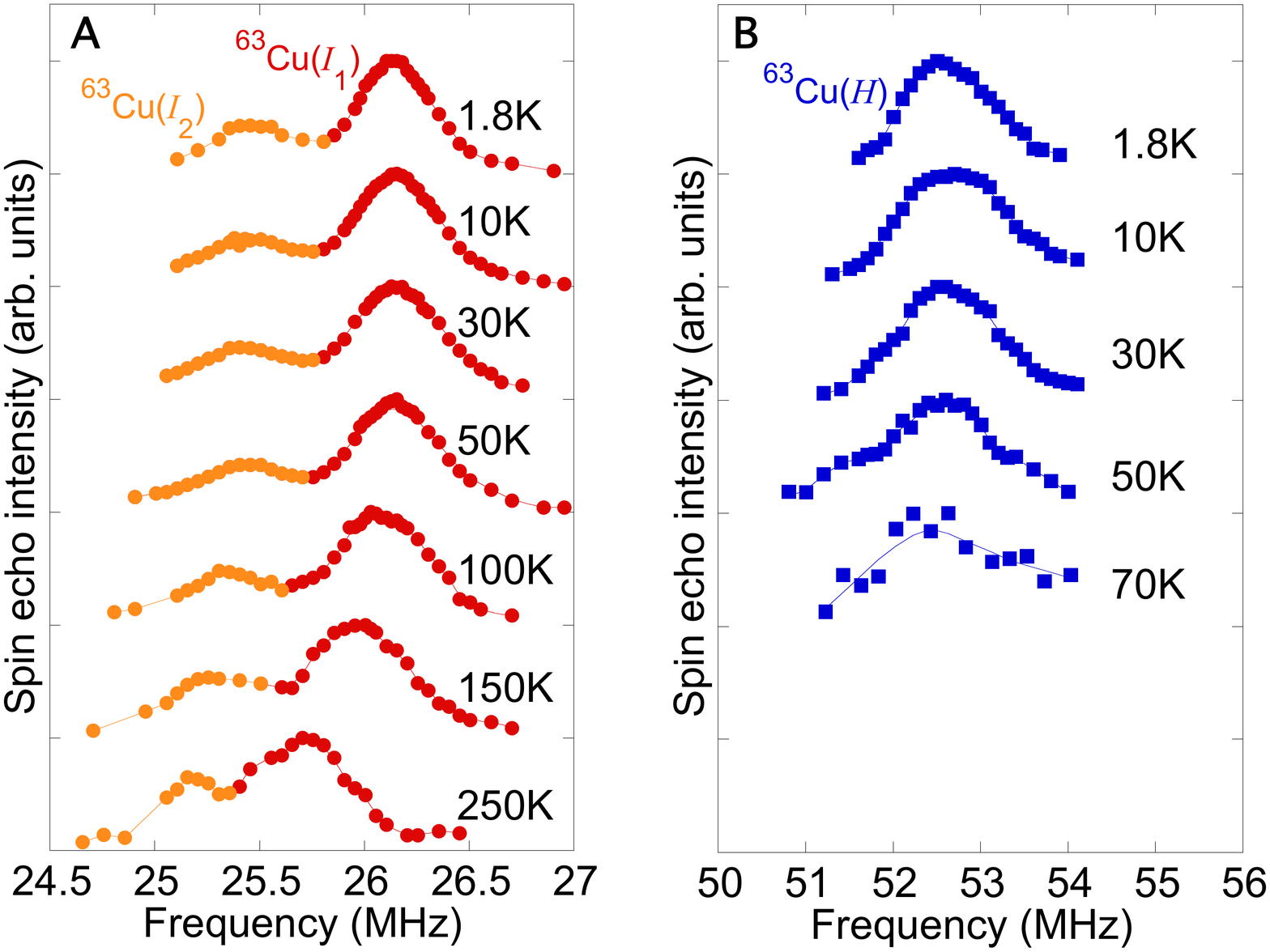}
		\includegraphics[width=3.2in]{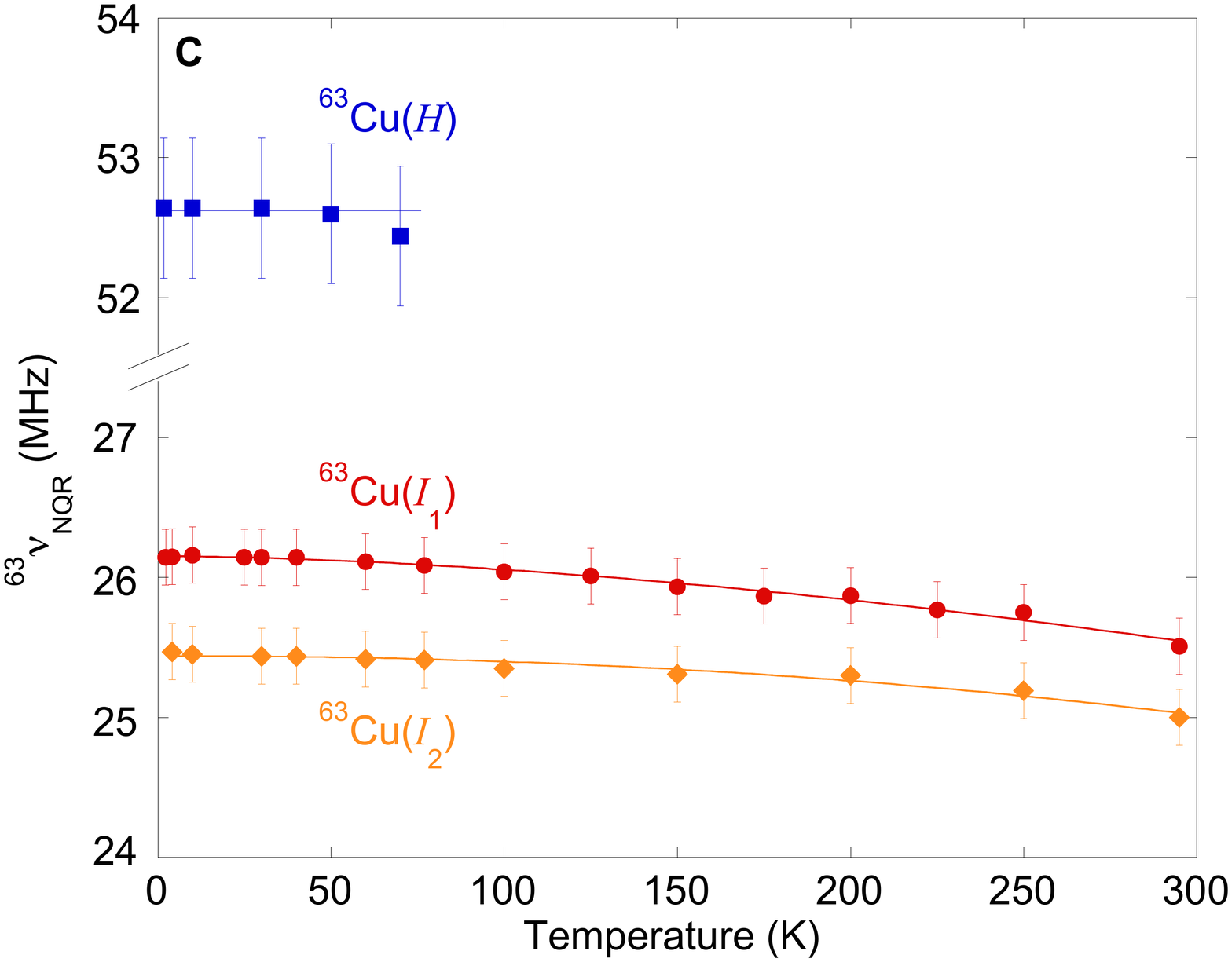}
		\caption{(A-B) $^{63}$Cu NQR peaks observed for the $^{63}$Cu(I$_{1,2}$) and $^{63}$Cu(H) sites at selected temperatures.  For clarity, the origin of the vertical axis is shifted at different temperatures, and the peak intensity is normalized to 1.  (C) Temperature dependence of $^{63}\nu_{NQR}$.  All solid lines are a guide for the eyes.}
		\label{fig:nuq}
	\end{center}
\end{figure}

\begin{figure}
	\begin{center}
		\includegraphics[width=3.2in]{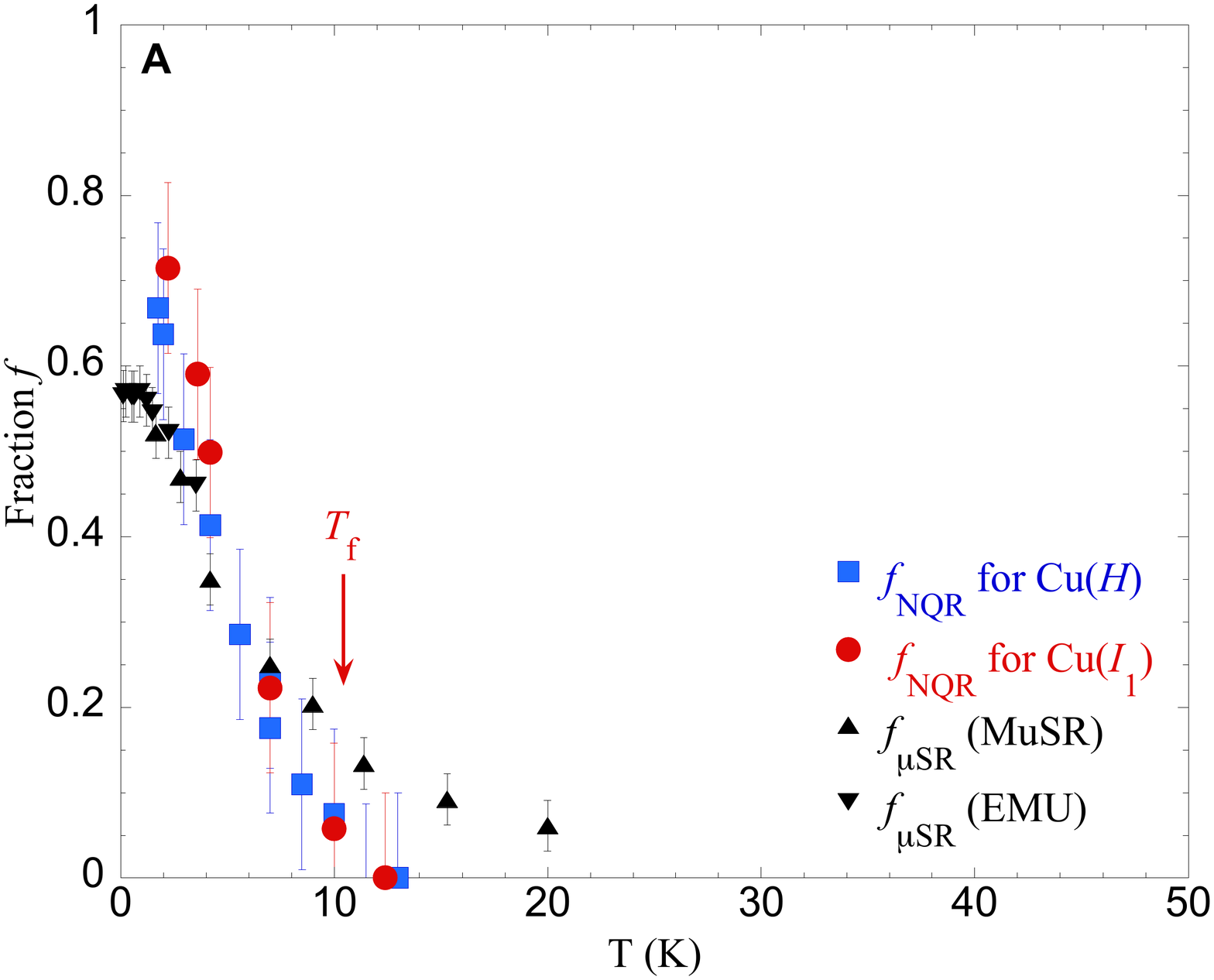}
		\includegraphics[width=3.2in]{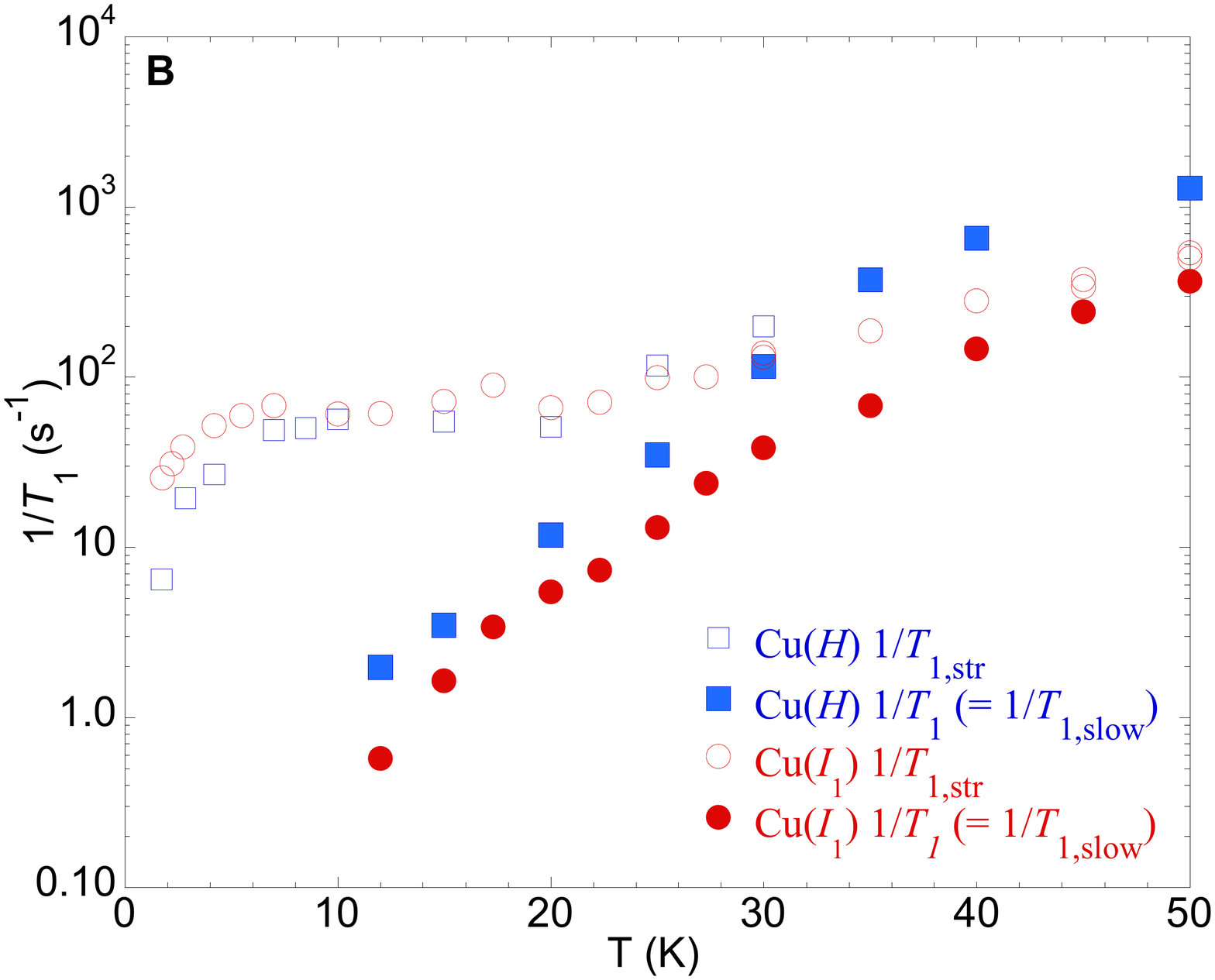}
		\caption{(A) The lost fraction $f_{NQR}$ of the $^{63}$Cu NQR signal intensity measured at the $^{63}$Cu(I$_1$) and $^{63}$Cu(H) sites at a fixed delay time $\tau = 10$~$\mu$s.  Also plotted are the fraction $f_{\mu SR}$ of the fast depolarizing component of the ${\mu}$SR signals, as measured by zero field ${\mu}$SR techniques using the MuSR (triangles) and EMU (downward triangles) spectrometers at the Rutherford Appleton Laboratory \cite{Kenney2018}.  (B) A precursor of the spin freezing observed above $T_f$ in the spin-lattice relaxation rate $1/T_1$.  Intrinsic $1/T_{1}$, obtained as the slower component $1/T_{1,slow}$ from the two component fit with eq.(A3) below $\sim 60$~K ({\it i.e.} $1/T_{1} = 1/T_{1,slow}$ below $\sim 60$~K), continues to slow down at low temperatures.  In contrast, the sample averaged behavior represented by $1/T_{1,str}$, estimated from the stretched fit of $M(t)$ with eq.(A2), levels off below $\sim$30 K.  See Appendix A for details of the fit.}
		\label{fig:lowT}
	\end{center}
\end{figure}

	We summarize the temperature dependence of $^{63}$Cu NQR lineshapes in Figs.\ 4A-B.  We observed no splitting or broadening of the NQR lineshapes due to static hyperfine magnetic fields arising from ordered magnetic moments, a typical NMR signature of antiferromagnetic long-range order (see Appendix C for magnetic NMR line broadening observed for Na$_2$IrO$_3$ below $T_N$).  The NQR frequency $^{63}\nu_{NQR}$ summarized in Fig. 4C reflects the local lattice and charge environment.  $^{63}\nu_{NQR}$ at $^{63}$Cu(I$_{1,2}$) decreases smoothly toward higher temperatures due to thermal expansion, without exhibiting evidence for a structural transition or dimerization.  We were unable to keep track of the $^{63}$Cu(H) NQR signals above $\sim$70 K due to extremely weak signal intensity caused by very fast NMR relaxation rates.  This is simply because the hyperfine coupling between $^{63}$Cu(H) nuclear spins with six nearest-neighbor Ir spins is stronger than that at the $^{63}$Cu(I$_{1,2}$) sites, as discussed in detail in Section III E below.  Generally, detection of the $^{63}$Cu NQR and NMR spin echo signals becomes difficult when the NMR spin-lattice relaxation rate $1/T_{1}$ reaches $\sim 10^{4}$ sec$^{-1}$ \cite{Imai2008}.   

	Although there is no evidence for magnetic long range order, we found that the integrated intensity of the $^{63}$Cu NQR signals begins to decrease gradually below $T_{f}\sim10$~ K.  This indicates that NMR relaxation rates become very fast in some domains.  In Fig. 5A, we summarize the temperature dependence of the lost fraction $f_{NQR}$ of the NQR spin echo intensity, measured at a constant pulse separation time $\tau =  10$ $\mu$s, down to 1.5~K.  $f_{NQR}$ shows qualitatively the same behavior as the volume fraction $f_{\mu SR}$ estimated by $\mu$SR experiments \cite{Kenney2018}, in which muons exhibit fast depolarization due to slow spin fluctuations in their vicinity \cite{Kenney2018,Choi2018}. Note that $f_{\mu SR}$ saturates at ~50\% below ~1 K.  ($f_{NQR}$ overestimates the volume fraction with slowed spin fluctuations, because the  distribution of the spin-spin relaxation time $T_2$ prevented us from accurately taking into account the contributions of nuclear spins with fast $T_2$.)  Combined with the fact that we found no evidence for divergent NMR relaxation rates for the observable NQR signals, we conclude that the loss of the $^{63}$Cu NQR signals are primarily due to slowing of defect spin fluctuations below $T_{f}$.  Since the $^{63}$Cu NQR lineshape for observable signals does not broaden even below $T_{f}$, the defect spins may be spatially confined in some domains, perhaps in the vicinity of stacking faults \cite{Kenney2018}, rather than uniformly distributed throughout the entire Ir$^{4+}$ honeycomb planes.

\subsection{Low energy Ir spin excitations in $B_{ext}=0$}
	We now turn attention to our central results on low-frequency Ir spin dynamics probed by $^{63}$Cu nuclear spin-lattice relaxation rate, $1/T_{1}$.  Quite generally, $1/T_{1}$ probes the low frequency component at the NQR frequency $\nu_{NQR}$ of the fluctuating hyperfine magnetic field $h_{perp} = A_{hf} S$ arising from the Ir spin $S$, i.e. $1/T_{1} = (\gamma_{n}^{2}/2) \int  <h_{perp}(t) h_{perp}(0)> exp(-2\pi i\nu_{NQR}t) dt$, where $\gamma_{n}/2\pi =$ 11.285 MHz/T is $^{63}$Cu nuclear gyromagnetic ratio, and {\it Perp} indicates the component orthogonal to the quantization axis of the observed nuclear spins.  The quantization axis of $^{63}$Cu nuclear spins is along the c-axis for our NQR measurements, and hence our NQR $1/T_{1}$ results probe the slow component at $^{63}\nu_{NQR}$ of Ir spin fluctuations within the ab-plane.  
	
	One can also write $1/T_{1} \propto \Sigma_{{\bf q}} |A_{hf}({\bf q})|^{2}S({\bf q},E_{n})$, where $A_{hf}({\bf q})$ is the wave-vector ${\bf q}$-dependent hyperfine form factor \cite{Millis1990}.  $S({\bf q},E_{n})$ is the dynamic spin structure factor of Ir spins at the very small excitation energy of $E_{n} = h\cdot~^{63}\nu_{NQR} \sim$~0.2 $\mu$eV, corresponding to the photon energy at the NQR frequency.  $1/T_{1}$ therefore probes the low energy sector of the ${\bf q}$-integrated spin excitation spectrum $S({\bf q},E_{n})$ at energy $E_n$, weighted by $|A_{hf}({\bf q})|^{2}$.  

\begin{figure}
	\begin{center}
		\vspace{0.08in}\includegraphics[width=3.3in]{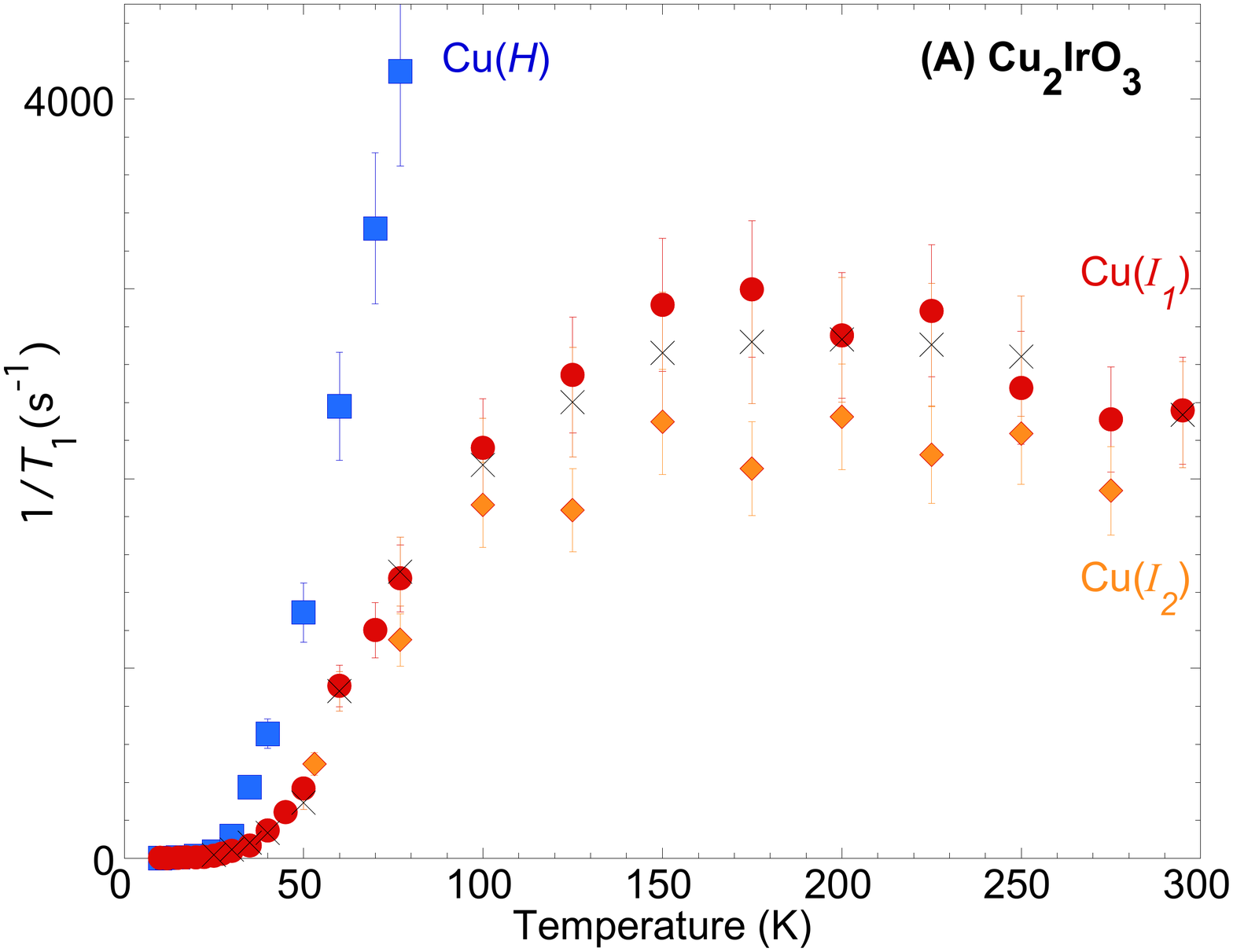}
		\includegraphics[width=3.3in]{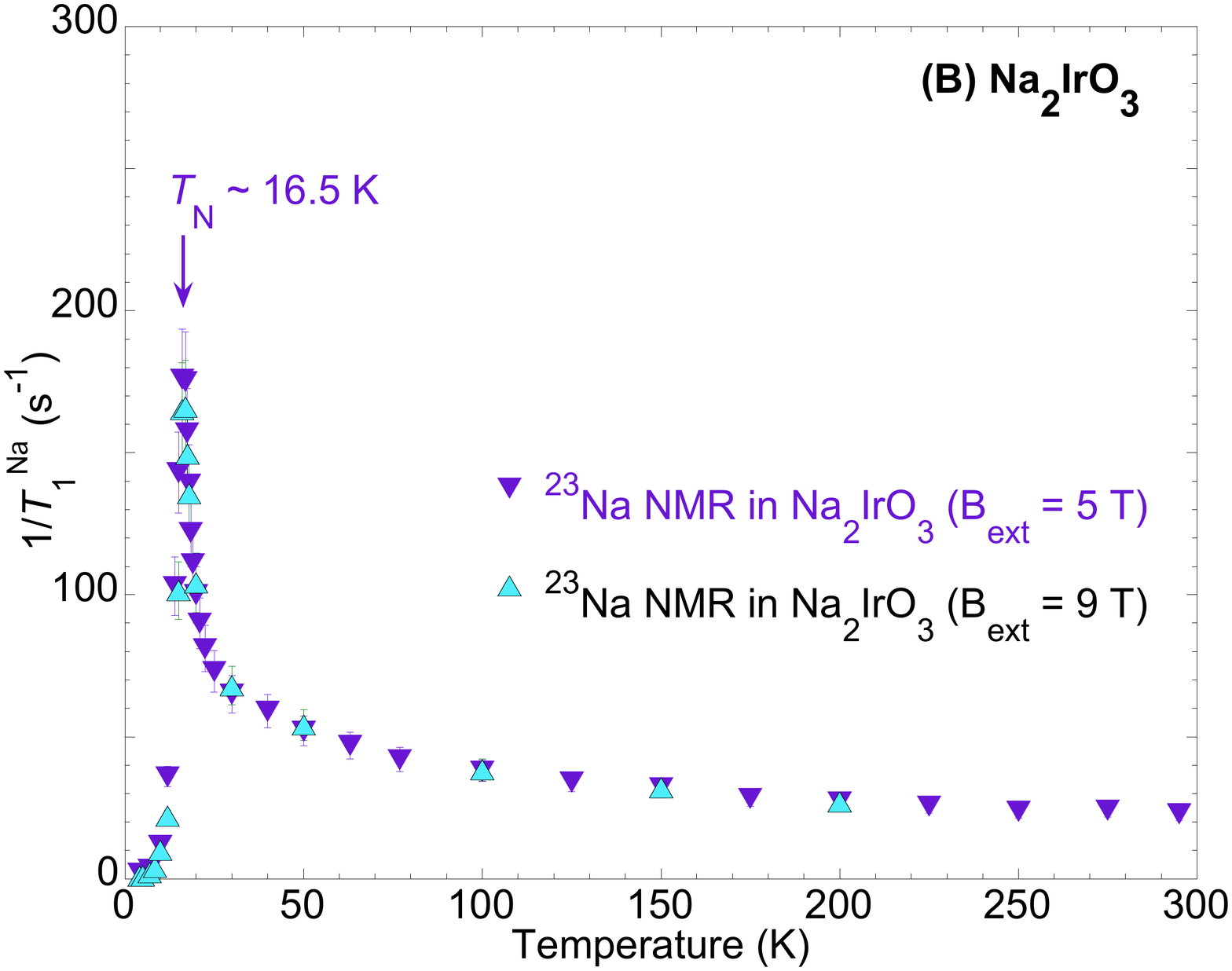}
		\caption{(A) Filled symbols: intrinsic $1/T_{1}$  measured in $B_{ext}=0$ with NQR at $^{63}$Cu(I$_1$), $^{63}$Cu(I$_2$) and $^{63}$Cu(H) sites of Cu$_2$IrO$_3$.  Note that $1/T_{1}$ plotted below $\sim 60$~K is the intrinsic, slow component estimated from the two component fit with eq.(A3) ({\it i.e.} $1/T_{1} = 1/T_{1,slow}$ below $\sim 60$~K). Also plotted with x symbols is $1/T_{1}$ measured for uni-axially aligned powder sample with NMR in $B_{ext}=9$~T applied along the ab-plane for the superposed peak of $^{63}$Cu(I$_1$) and $^{63}$Cu(I$_2$) sites.  (B) $1/T_{1}^{Na}$ observed for the inter-planar $^{23}$Na(I) sites in a powder sample of Na$_2$IrO$_3$ in $B_{ext}=5$~T and 9 T.}
		\label{fig:T1}
	\end{center}
\end{figure}

\begin{figure}
	\begin{center}
		\hspace*{0.15in}\includegraphics[width=3.4in]{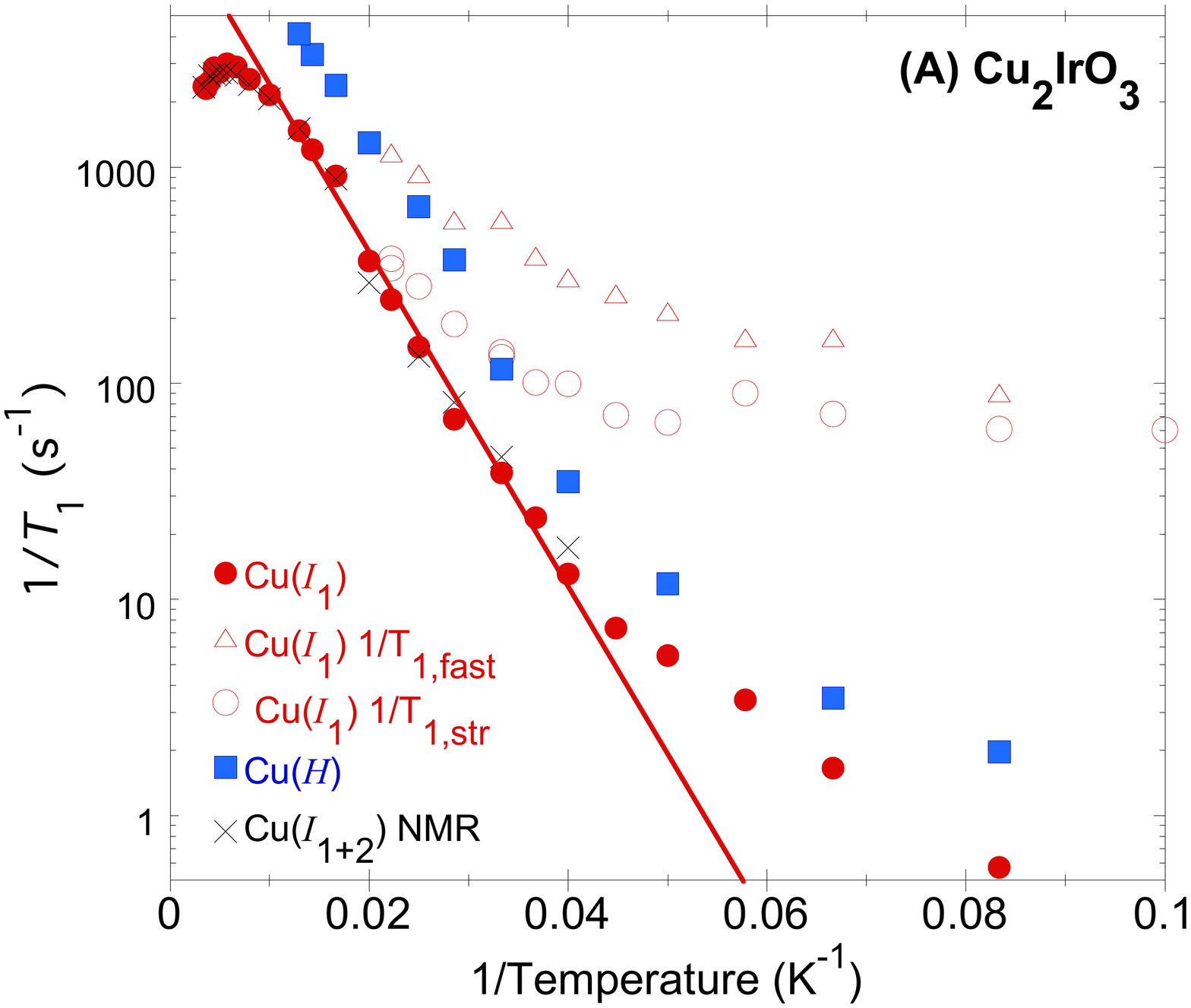}
		\includegraphics[width=3.7in]{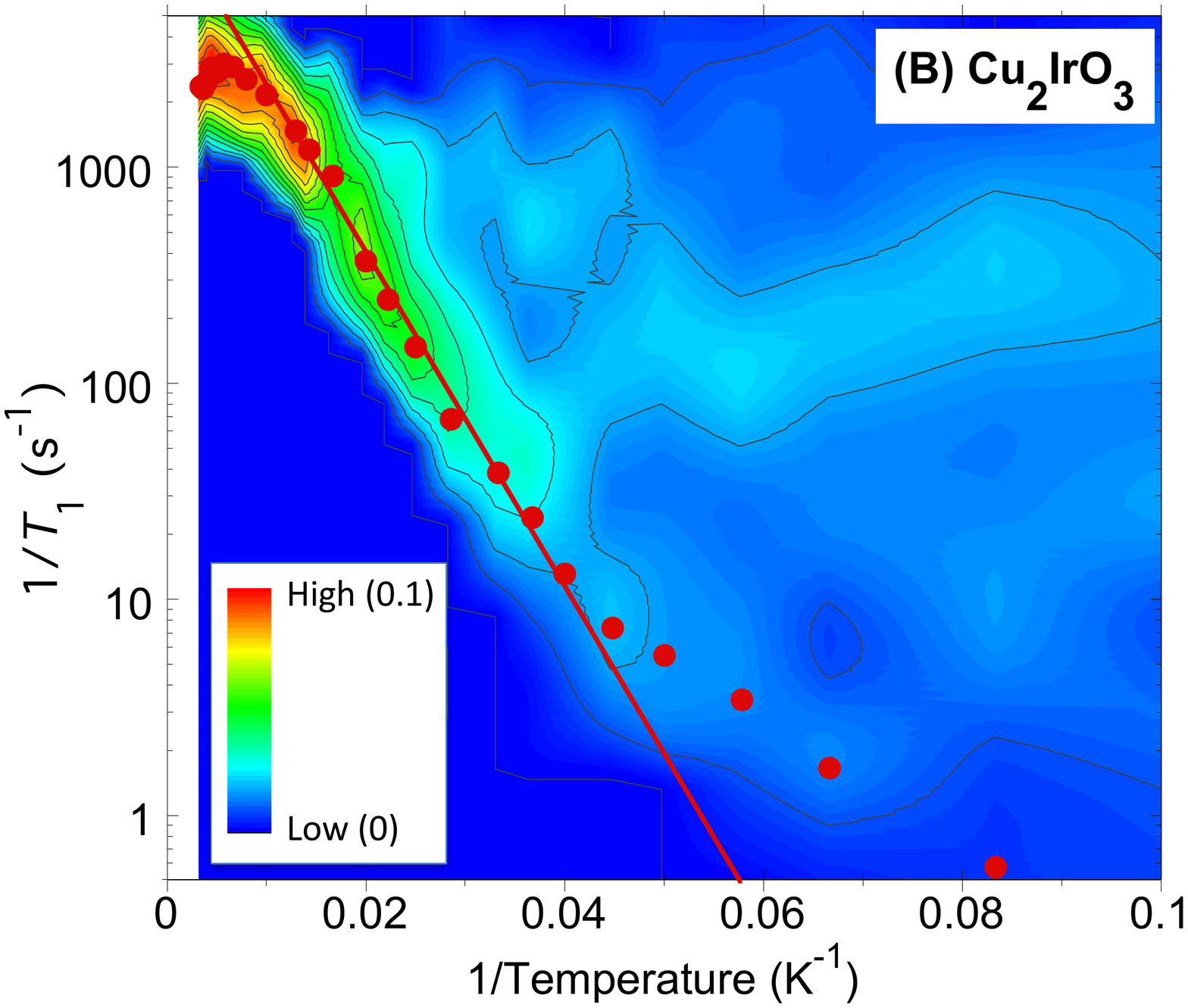}
		\caption{(A) The same NQR and NMR $1/T_{1}$ results as in Fig.~6A, plotted as a function of inverse temperature $1/T$ at the $^{63}$Cu(I$_1$) and $^{63}$Cu(H) sites (filled symbols).  The solid line shows an activation behavior, $1/T_{1} \sim exp(- \Delta /k_{B}T)$ with a gap, $\Delta/k_{B}=175 \pm 30$~K.  Open triangles represent the fast component $1/T_{1,fast}$ in the two-component fit with eq.(A3), while open bullets show the averaged behavior $1/T_{1,str}$ as estimated from the stretched single exponential fit with eq.(A2), both at the $^{63}$Cu(I$_1$) sites.  
(B) The color contour plot of the temperature dependence of the probability distribution $P(1/T_{1})$ of $1/T_{1}$ at $^{63}$Cu(I$_{1}$) sites measured with NQR, as determined by inverse Laplace transform of $M(t)$.  We overlaid the NQR $1/T_{1}$ at $^{63}$Cu(I$_1$) sites and its activation behavior, $1/T_{1} \sim exp(- \Delta /k_{B}T)$, from (A).
		}
		\label{fig:logT1}
	\end{center}
\end{figure}

Summarized in Figure 6A is $1/T_{1}$ at the $^{63}$Cu(H), $^{63}$Cu(I$_1$), and $^{63}$Cu(I$_{2}$) sites measured at the peak frequency of the NQR lineshapes.  The hyperfine coupling $A_{hf}$ of Ir spins with the $^{63}$Cu(H) nuclear spins is stronger than with Cu(I$_{1,2}$) sites due to their spatial proximity, and hence the magnitude of $1/T_{1}$ at the $^{63}$Cu(H) sites is larger by a factor of $\sim 3$.  As noted above, this is the underlying reason why the $^{63}$Cu(H) NQR signal detection becomes difficult above $\sim$70 K.  Otherwise, the temperature dependence of $1/T_{1}$ is qualitatively the same between the $^{63}$Cu(H), $^{63}$Cu(I$_1$), and $^{63}$Cu(I$_2$) sites. 

$1/T_{1}$ grows slightly below 300 K, a typical signature of the gradual development of short-range spin-spin correlations, in the present case between Ir spins.  It is followed by an onset of a dramatic suppression of $1/T_{1}$ below $\sim$150 K.  A semi-logarithmic plot in Figure 7A indicates that $1/T_{1}$ follows an activation behavior over two decades, from $\sim$100 K to $\sim$20 K, due to a complete suppression of the low frequency component of the fluctuating hyperfine magnetic fields $h_{perp}$.  By fitting the result to an activation form $1/T_{1} \sim exp(- \Delta/k_{B}T)$, we estimate the gap, $\Delta/k_{B} = 175 \pm 30$ K ($\Delta \sim15$ meV).  

We note that $1/T_{1}$ develops a distribution below $\sim$60 K, and hence the $1/T_1$ results plotted below $\sim 60$~K with filled symbols in Figs.\ 5B, 6A and 7 represent the intrinsic slow component (i.e. $1/T_{1,slow}$) estimated from the two component fit with eq.\ (A3), as explained in detail in Appendix A.  We emphasize, however, that our key finding of the activation behavior in Figs.\ 7A does not depend on the details of how we estimate $1/T_{1}$ from the nuclear spin recovery curve $M(t)$.  The proof comes from the distribution function $P(1/T_{1})$ of $1/T_{1}$ directly calculated from the inverse Laplace transform of the nuclear magnetization $M(t)$ \cite{Johnston2005,Singer2018}, as explained in the next section C. 
	
Also plotted in Fig.\ 5B and 7 using open bullets are $1/T_{1,str}$ obtained from the phenomenological stretched fit of $M(t)$ with eq.\ (A2) \cite{Thayamballi1980, Itoh1986}.  As explained in section III-C, $1/T_{1,str}$ is a good measure of the center of gravity of the distributed $1/T_1$.   As the influence of inhomogeneous spin fluctuations from defects grow near $T_f$, the aforementioned activation behavior of the intrinsic slow component of $1/T_{1}$ is terminated by the precursor of spin freezing, which also results in a shoulder of $1/T_{1,str}$ below $\sim$30 K as shown in Fig.\ 5B.  This is followed by a gradual loss of $^{63,65}$Cu NQR signal intensity below $T_{f} \sim 10$~K in Fig. 5A.  The fact that a half of the NQR signal remain observable without exhibiting any magnetic anomalies (such as the line broadening, divergence of $1/T_{1}$, or signal loss) indicates that a majority of Ir spins themselves are not slowing down toward eventual long range order or freezing.  Instead, it is the inhomogeneously distributed defects (perhaps accompanied by minority Ir spins adjacent to them) that are freezing.  This is consistent with $\mu$SR observation that Ir spins remain dynamic even below $T_f$ \cite{Kenney2018, Choi2018}, and also with a very wide distribution of $1/T_1$ near $T_f$ as determined by inverse Laplace transform analysis in section III-C.  Whether these defects originate from randomly distributed Cu$^{2+}$ ions and/or clusters around the stacking faults encompassing several unit cells remains to be seen. 
	
\subsection{Inverse Laplace transform of $M(t)$ for direct estimation of the $1/T_1$ distribution}
A more precise, model-independent way to investigate the spin dynamics with $1/T_1$ under the presence of distributions in the relaxation mechanisms is to take the inverse Laplace transform of $M(t)$, and directly deduce the distribution function $P(1/T_{1})$ of $1/T_{1}$ \cite{Johnston2005,Singer2018},  
\begin{equation}
M(t) = \sum_{i}\left[M_{o}-A \cdot exp(-3t/T_{1,i})\right] P(1/T_{1,i}),	
\end{equation} 	
where we normalize the overall probability to 1, $\sum_{i}P(1/T_{1,i}) = 1$.  We numerically inverted $M(t)$ \cite{Singer2018} utilizing Tikhonov regularization (i.e. a smoothing factor) \cite{Venkataramanan2002, Song2002}, and deduced the distribution function $P(1/T_{1})$ as summarized in Fig.\ 8 at representative temperatures.  

\begin{figure}
	\begin{center}
		\includegraphics[width=3in]{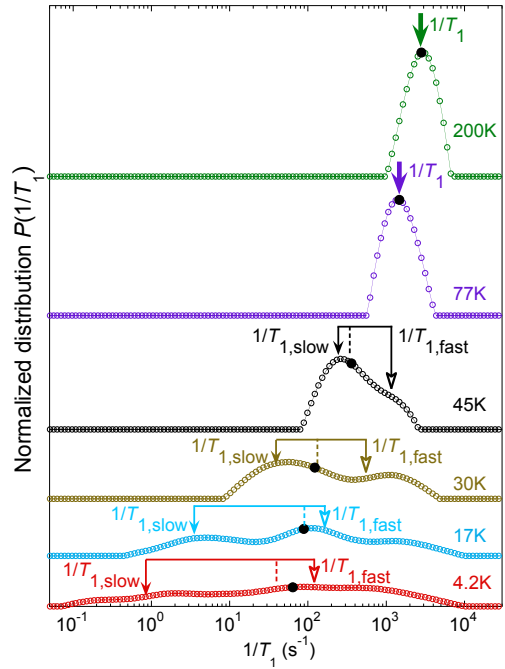}
		\caption{Distribution function $P(1/T_{1})$ of $1/T_1$ at representative temperatures directly computed from the recovery data $M(t)$ based on the inverse Laplace transform \cite{Singer2018}.  For clarity, the origin is shifted vertically except for 4.2~K.  $P(1/T_{1})$ is centered around a single value of $1/T_1$ above ~60 K; the peak location at 77 K and 200 K agrees very well with $1/T_1$ estimated from the single component fit with eq. (A1), as marked by thick arrows.  This one peak distribution of $P(1/T_{1})$ breaks down at lower temperatures.  The filled and open arrows at and below 45 K mark $1/T_{1,slow}$ and $1/T_{1,fast}$ obtained from the two component fit with eq.(A3), respectively.  The dashed lines mark $1/T_{1,str}$ estimated from eq.(A2), which keeps track of the center of gravity of $P(1/T_{1})$ marked by black filled bullets. }
		\label{fig:ILT}
	\end{center}
\end{figure}

We do not presume any phenomenological functional forms for $M(t)$, such as the existence of two separate components in eq.(A3).  Nonetheless, our $P(1/T_{1})$ results indicate that a new small side peak begins to emerge with fast values of $1/T_{1}$ below $\sim$60 K.  The emergence of the two peak structure in $P(1/T_{1})$ indicates that the intrinsic, slow $T_1$ process begins to be short circuited by the faster component(s) in the vicinity of defect spins as the intrinsic $T_{1}$ slows down.  Since the influence of the defect-induced fast relaxation depends on the distance with observed nuclear spins, defect induced fast component in $1/T_{1}$ generally has a broad distribution.

To show the key features of $P(1/T_{1})$ visually, we plotted $P(1/T_{1})$ in a color contour map in Fig.\ 7B.  We can identify the existence of the well-defined, slower component of $1/T_{1}$ in the ridge structure (colored in red or green) that continues from $\sim$60 K down to $\sim$15 K (i.e. $1/T = 0.015$ to 0.07).   On the other hand, the defect induced fast contribution to $1/T_{1}$ develops a wide distribution centered around $\sim 150$~s$^{-1}$, and remains roughly constant, see the light blue horizontal section starting from $1/T \sim 0.04$ to 0.1 in Fig.\ 7B.  Interestingly, recent theoretical calculations of $1/T_{1}$ for Kitaev QSL predict that the presence of bond-disorder leads to a constant $1/T_{1}$ with a broad distribution \cite{Knolle2019}. We emphasize that our observation of the activation behavior for the intrinsic slow component and the roughly constant behavior induced by defects, both nicely captured in Fig.\ 7B, does not depend on fitting procedures of $M(t)$, such as the stretched nature of the relaxation function in eq.(A2) or the presence of two components in eq.(A3).   

Also notice that $1/T_{1,slow}$ and $1/T_{1,fast}$ estimated from the two component fit (marked by thin filled and thin open arrows, respectively, in Fig.\ 8) agree fairly well with the two peaks observed for $P(1/T_{1})$ in a wide temperature range except near and below $T_f$, where the two component fit itself becomes invalid.  This justifies the two component fit with eq.(A3) used above $T_f$ to generate intrinsic $1/T_1$ as $1/T_{1,slow}$ in Figs.\ 5B, 6A and 7.  Interestingly, $1/T_{1,str}$ estimated using eq.(A2) (marked by vertical dashed lines in Fig.\ 8) agrees well with  the center of gravity of the distribution $P(1/T_{1})$ (marked by black filled bullets).  This means that $1/T_{1,str}$ probes only the spatially averaged behavior of the entire sample, and does not always represent the intrinsic behavior of $1/T_1$.  

\subsection{Comparison with low energy Ir spin excitations of Na$_2$IrO$_3$, $\alpha$-RuCl$_3$, and H$_3$LiIr$_2$O$_6$}
	It is illustrative to compare the $1/T_{1}$ results of Cu$_2$IrO$_3$ with those observed for other Kitaev candidate materials. In Fig. 6B, we summarize $1/T_{1}^{Na}$ measured for the inter-planar $^{23}$Na(I) sites of Na$_2$IrO$_3$ at the central peak of the powder $^{23}$Na NMR lineshape.  We refer readers to Appendix C for the details of $^{23}$Na NMR measurements.  Simutis et al. also reported analogous $1/T_{1}^{Na}$ data in ambient and applied pressure \cite{Simutis2018}.  Their results, reported in a limited temperature range up to 25~K, are in good agreement with ours. The overall magnitude of $1/T_{1}^{Na}$ is nearly two orders of magnitude slower than at the Cu(I$_{1,2}$) sites in Fig.\ 6A, because the hyperfine coupling $A_{hf}$ is much smaller at the $^{23}$Na sites, as evidenced by the $^{23}$Na NMR Knight shift results (see Fig.\ 11B in Section III E).  
	
Below 300 K down to $\sim$25 K, $1/T_{1}^{Na}$ gradually increases as the short-range Ir-Ir spin correlations grow, without exhibiting a downturn observed for Cu$_2$IrO$_3$.  The critical slowing down of Ir spin fluctuations below $\sim$25 K toward the three dimensional zig-zag antiferromagnetic long range order results in a sharp divergent behavior of $1/T_{1}^{Na}$ at $T_{N} = 16.5 \pm 0.5$~ K. This is consistent with the fact that the hyperfine form factor does not cancel at the $^{23}$Na(I) sites for the zig-zag order pattern  (schematically shown in Fig.\ 2B).    Analogous divergent behavior is commonly observed for conventional antiferromagnets, such as CuO \cite{Itoh1990}.  
	  
Below $T_N$, $1/T_{1}^{Na}$ is quickly suppressed, accompanied by dramatic broadening of the $^{23}$Na powder NMR lineshape due to the static hyperfine magnetic fields from zig-zag ordered Ir spins (see Fig.\ 17 in Appendix C as well as \cite{Simutis2018}).  In general, $1/T_{1}$ in the antiferromagnetically ordered state below $T_N$ is dominated by two- or three-magnon Raman process, resulting in suppression of $1/T_{1}$ obeying a power law; it is generally followed by an activation behavior $1/T_{1} \sim T^{2} \cdot exp(-\Delta_{magnon}/k_{B}T)$ due to the anisotropy gap $\Delta_{magnon}$ for magnons at lower temperatures below $\Delta_{magnon}/k_{B}$ \cite{Beeman1968, Tsuda1992}.  While our $1/T_{1}^{Na}$ results were measured for broad and superposed $^{23}$Na NMR lines and do not allow us to conduct detailed analysis below $T_N$, the observed behavior of $1/T_{1}^{Na}$ below $T_N$ is similar to the conventional behavior expected for three-dimensional or quasi-two dimensional antiferromagnets, such as CuO and YBa$_2$Cu$_3$O$_6$ \cite{Tsuda1992}.  
	
Our conventional findings on Na$_2$IrO$_3$ are no surprise, in view of the fact that earlier inelastic neutron scattering measurements established that the low energy sector of the spin excitation spectrum in Na$_2$IrO$_3$  is dominated by magnons that are not native to pure Kitaev Hamiltonian in eq.(1) \cite{Choi2012}. In this context, it is also interesting to compare our results with those previously reported for $\alpha$-RuCl$_3$ \cite{Baek2017,Zheng2017,Jansa2018}, because magnons also dominate its spin excitation spectrum in the low field regime \cite{Banerjee2016,Banerjee2017Science,Banerjee2018}.  Not surprisingly, $1/T_{1}$ observed at the $^{35}$Cl site of $\alpha$-RuCl$_3$ in the low field regime $B_{ext} = 2.35$~T with antiferromagnetic ground state \cite{Jansa2018} is similar to the conventional behavior observed here for Na$_2$IrO$_3$.  Our NQR results for Cu$_2$IrO$_3$ in Figs.\ 6A and 7 are fundamentally different.   

In the case of $\alpha$-RuCl$_3$, application of a modest magnetic field above $\sim 7$~T drives the antiferromagnetic ground state to a field-induced paramagnetic state with a gap \cite{Baek2017,Zheng2017,Jansa2018}.  This is often attributed to a spin liquid state.  The report of fractionalized thermal Hall effect \cite{Kasahara2018} in the intermediate field regime of  $\alpha$-RuCl$_3$ as well as peculiar magnetic field effect on magnon dispersion \cite{Banerjee2018} certainly raises the hope that chiral spin-liquid with Majorana edge modes may be  induced by a magnetic field, but even the classical spin wave theory can semi-quantitatively account for the former \cite{Chern2019}.   We also caution that application of an external magnetic field could fundamentally alter the nature of spin excitations of quantum spin systems.  For example, $B_{ext}$ exceeding 5.3 T applied to the transverse field Ising chains realized in CoNb$_2$O$_6$ suppresses a magnetic long-range order and induces a spin excitation gap that manifests itself in the $^{59}$Co NMR properties \cite{Kinross2014}.  The nature of the field-induced gap and its potential relation to Kitaev physics has been at the center of recent intense debate \cite{Baek2017,Zheng2017,Jansa2018,Banerjee2018,Kasahara2018}, and deserves further careful examinations.  

 Finally, we compare Cu$_2$IrO$_3$ and H$_3$LiIr$_2$O$_6$ \cite{Kitagawa2018}; both of these Kitaev candidate materials remain paramagnetic without undergoing magnetic long-range order.  Interestingly, $1/T_{1}$ observed at the intra-layer $^{7}$Li sites of H$_3$LiIr$_2$O$_6$ shows qualitatively the same behavior as the intra-layer $^{63}$Cu(H) as well as inter-layer $^{63}$Cu(I$_{1}$)  sites from 300~K down to $\sim$50~K, where the intrinsic behavior of $^{7}$Li $1/T_1$ appears to be taken over by large defect induced contributions.  On the other hand, $1/T_{1}$ observed at the inter-layer $^{1}$H sites in H$_3$LiIr$_2$O$_6$, which corresponds to the $^{63}$Cu(I) sites in the present case, shows no hint of a gapped behavior at low temperatures.  This apparent discrepancy from the $^{7}$Li or our  $^{63}$Cu results, however, does not necessarily mean that H$_3$LiIr$_2$O$_6$ does not have analogous activation behavior of $1/T_{1}$.  It is well-known that $^{1}$H NMR $1/T_{1}$ results in herbertsmithite kagome antiferromagnet ZnCu$_3$(OH)$_6$Cl$_2$ \cite{Imai2008} are dominated by defect spins, especially in low magnetic fields, and do not resemble the intrinsic $1/T_{1}$ results observed at $^{63}$Cu \cite{Imai2008} or $^{17}$O \cite{Fu2015} sites.  This is simply because Fermi's contact hyperfine coupling of $^{1}$H nuclear spin through the 1s orbital is generally very weak, and hence $1/T_{1}$ measured at $^{1}$H sites is comparatively more sensitive to defect induced spin fluctuations.  We note that the $^{1}$H as well as $^{7}$Li  NMR linewidth in H$_3$LiIr$_2$O$_6$ is greater than the small NMR frequency shift ({\it i.e.} Knight shift is very small).  This suggests that a large fraction of nuclear spins have stronger hyperfine couplings with defect spins than with nearby Ir$^{4+}$ spins with intrinsic Kitaev properties, which explains why $^{7}$Li and $^{1}$H NMR $1/T_1$ results are overwhelmed by defect contributions.  In the case of Cu$_2$IrO$_3$, this does not happen, because the transferred hyperfine couplings of $^{63}$Cu nuclear spins through the 4s orbital \cite{Mila-Rice} with surrounding Ir sites are larger by an order of magnitude, as demonstrated in the next section. 

\subsection{NMR measurements in applied magnetic field}
	We also conducted high-field NMR measurements using a uni-axially aligned powder sample of Cu$_2$IrO$_3$ cured in a glue (stycast 1266) in the presence of $B_{ext}=9$~T.  This is a standard trick used in powder NMR experiments, and creates a pseudo single crystal by taking advantage of the anisotropy of magnetization in high magnetic fields \cite{Takigawa1989}.  We summarize representative $^{63}$Cu NMR lineshapes observed for the nuclear spin $I_{z} = +1/2$ to $-1/2$ central transition of the aligned powder sample in $B_{ext}=9$~T applied along the aligned c- and ab-axis in Fig. 9B and C, respectively.  See Fig.\ 15 in Appendix B for the field sweep NMR spectra covering the satellite transitions between the nuclear spin $I_{z} = \pm1/2$ to $\pm3/2$ states.  From the comparison with the NMR lineshapes for unaligned powder in Fig. 9A, we estimate the fraction of the uni-axially aligned powder as approximately 50\%; the rest of the powder remains randomly oriented. 
	
	We could assign the $^{63,65}$Cu NMR peaks and shoulders arising from Cu(I$_{1,2}$) sites in aligned and unaligned portions of the sample, as marked by arrows, + and x symbols in Fig. 9A-C.  However, identification of the NMR signals from less abundant $^{63}$Cu(H) sites proved to be a major challenge, because the $^{63,65}$Cu(H) central transition is shifted by the second and higher order effects caused by the extremely large nuclear quadrupole frequency ($^{63}\nu_{NQR}\sim 52$~MHz), and superposed by much stronger central and satellite transitions of the $^{63,65}$Cu(I$_{1,2}$) sites.  Moreover, since $^{63}\nu_{NQR}$ at the $^{63}$Cu(H) sites happens to be almost exactly twice larger than $^{63}\nu_{NQR}$ at the $^{63}$Cu(I$_{1,2}$) sites, the small $I_{z} = \pm 3/2$ to $\pm1/2$ satellite peaks of the $^{63}$Cu(H) sites are accidentally superposed by the satellite structures arising from the $^{63}$Cu(I$_{1,2}$) sites.  To make matters worse, defect spins polarized by $B_{ext}$ broaden all the NMR lines below $\sim$50 K, where $^{63}$Cu(H) NQR signal was readily observable.  Nonetheless, we were able to resolve the $^{63}$Cu(H) NMR central transition peak below $\sim$120 K in the $B_{ext}~||$ c geometry, as marked by blue arrows in Fig. 9B.  
	
	The resonance frequency of the $^{63}$Cu(I$_{1,2}$) and $^{63}$Cu(H) peak for $B_{ext}~||$~c in Fig.\ 9B is shifted from the bare Zeeman frequency $^{63}\nu_{o}$ marked by the vertical dashed line due to the paramagnetic NMR Knight shift, $^{63}K= N_{nn}A_{hf}\chi_{spin}/N_{A}\mu_{B}$, where $N_{nn}$ is the number of the nearest neighbor Ir sites that induce transferred hyperfine magnetic fields at Cu nuclear spins, $N_A$ is Avogadro's number, $\mu_B$ is Bohr magneton; $\chi_{spin}$ is the spin contribution to the bulk susceptibility $\chi = \chi_{spin} + \chi_{dia} + \chi_{vv}$, where the diamagnetic term $\chi_{dia} \sim -0.09 \times 10^{-3}$ emu/mol from the standard table, and $\chi_{vv} \sim 0.16 \times 10^{-3}$ emu/mol \cite{Chaloupka2013}.  $^{63}$Cu(I$_{1,2}$) sites are bonded with two O sites above and below (see Fig. 1E), each of which bonds with two Ir sites, resulting in $N_{nn} = 4$.  On the other hand, $N_{nn} = 6$ for $^{63}$Cu(H), as may be seen in Fig. 1D.  
	
In the case of the $^{63}$Cu(I$_{1,2}$) peak in $B_{ext}~||$~ab, the apparent Knight shift $^{63}K_{apparent}$ is also affected by the second order contribution of the nuclear quadrupole interaction $\Delta \nu_{Q}^{(2)}$, which is proportional to $1/(\gamma_{n}B_{ext})^{2}$.  We therefore measured $^{63}K_{apparent}$ at different magnetic field values, linearly extrapolated the results to the high field limit of $1/(\gamma_{n}B_{ext})^{2} = 0$ \cite{Takigawa1989}, and estimated $^{63}K$ as shown in Fig.\ 10.  On the other hand, within experimental uncertainties, we found no dependence of $^{63}K_{apparent}$ on $1/(\gamma_{n}B_{ext})^{2}$ for both $^{63}$Cu(I$_{1,2}$) and $^{63}$Cu(H) sites in the $B_{ext}~||$~c geometry.  This implies that the nuclear quadrupole effects $\Delta \nu_{Q}^{(2)}$ play no significant roles in the NMR frequency shift for $B_{ext}~||$~c, because the main principal axis of the EFG tensor is along the c-axis and the asymmetry parameter $\eta$ of the EFG tensor is negligible ($\eta \sim 0$).  This is consistent with the local symmetry of these sites.  
	
	We summarize the temperature dependence of $^{63}K$ in Fig.\ 11A in comparison to $\chi_{spin}$ measured in 5 T ($\chi_{spin}$ does not depend on magnetic field above $\sim 30$~K \cite{Abramchuk2017}).  $^{63}K$ increases with decreasing temperature, and saturates below $\sim$30 K.  Interestingly, $^{63}K$ observed for Cu$_2$IrO$_3$ shows qualitatively the same site and temperature dependences as the $^{23}$Na NMR shift $^{23}K$ observed for Na$_2$IrO$_3$ in the paramagnetic state above $T_{N} \sim 16.5$~K, as shown in Fig.\ 11B.  Noting that $^{63}K$ probes local spin susceptibility rather than the bulk averaged result, our $^{63}K$ results indicate that the upturn of the bulk $\chi$ data observed below $\sim$30 K does not reflect the intrinsic behavior of the Ir honeycomb planes, but should be attributed to defect spins.  It is the enhanced local spin susceptibility of the defect spins that inhomogeneously broadens our high field NMR peaks at low temperatures.  

We plotted $^{63}K$ vs. $\chi_{spin}$ choosing temperature as the implicit parameter in Fig.\ 12, and estimated the hyperfine coupling from the slope of the linear fit as $A_{hf} = (N_{A} \mu_{B}/N_{nn}) \cdot d(^{63}K)/d \chi_{spin} = 2.3$ and 7.6 kOe/$\mu_B$ for $^{63}$Cu(I$_{1,2}$) in the $B_{ext}~||$~c and $||$~ab geometry, respectively, and $A_{hf} = 4.4$ kOe/$\mu_B$ for $^{63}$Cu(H) in $B_{ext}~||$~c.  We were unable to determine $A_{hf}$ at the $^{63}$Cu(H) sites along the ab direction, because the small central peak is buried underneath the $^{65}$Cu(I$_{1,2}$) signals due to extremely large $\Delta \nu_{Q}^{(2)}$ as explained earlier.     	

\begin{figure}
	\begin{center}
		\includegraphics[width=3.2in]{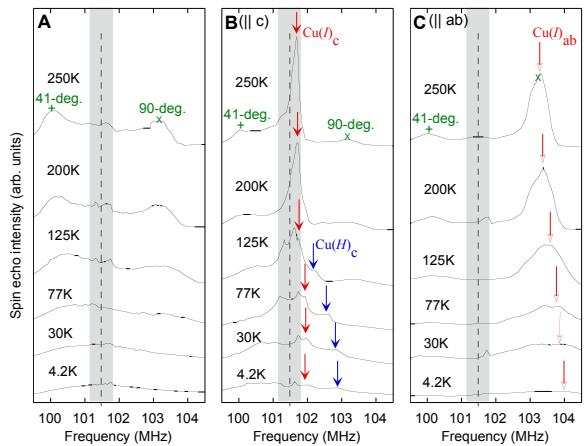}
		\caption{(A) Representative $^{63}$Cu NMR lineshapes observed for an unaligned powder sample in an external magnetic field $B_{ext}= 9$~T.  Green x and + mark the peaks corresponding to the particles in which $B_{ext}$ is 90-degrees (i.e. ab-plane) and 41-degrees tilted from the main-principle axis of the electric field gradient tensor along the c-axis.  The grey shaded region is superposed by weak $^{23}$Na NMR signals centered at $\sim$101.3 MHz arising from a small amount of impurity phase and the background $^{63}$Cu NMR signals at $\sim$101.7 MHz from the metallic parts used in our cryogenic NMR probe. Dashed vertical line shows the unshifted $^{63}$Cu NMR frequency $^{63}\nu_{o} = \gamma_{n}B_{ext}$.  (B) Representative $^{63}$Cu NMR lineshapes observed for a uni-axially aligned powder sample in $B_{ext}=9$~T applied along the aligned c-axis.  Red filled arrows mark the $^{63}$Cu(I) peak, while the blue filled arrows mark the $^{63}$Cu(H) peak, both in the $B_{ext}~||$~c geometry.  (C) Representative $^{63}$Cu NMR lineshapes observed for a uni-axially aligned powder sample in $B_{ext}=9$~T applied along the aligned ab-plane.  Red open arrows mark the $^{63}$Cu(I) peak in $B_{ext}~||$~ab geometry.}
		\label{fig:NMR}
	\end{center}
\end{figure}

\begin{figure}
	\begin{center}
		\includegraphics[width=3.2in]{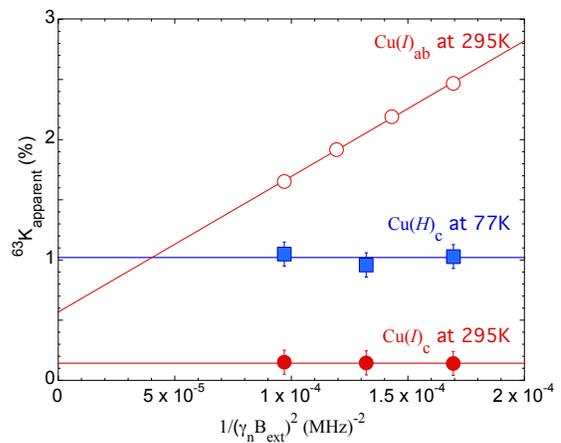}
		\caption{Separation of the NMR Knight shift $^{63}K$ contribution from the second order quadrupole contribution $\Delta \nu_{Q}^{(2)}$ based on the measurements of $^{63}K_{apparent}$ in several magnetic fields.  The linear extrapolation of $^{63}K_{apparent}$ to the limit of $1/(\gamma_{n}B_{ext})^{2} = 0$ gives $^{63}K$.}
		\label{fig:Kapparent}
	\end{center}
\end{figure}

\begin{figure}
	\begin{center}
		\hspace*{0.07in}\includegraphics[width=3.2in]{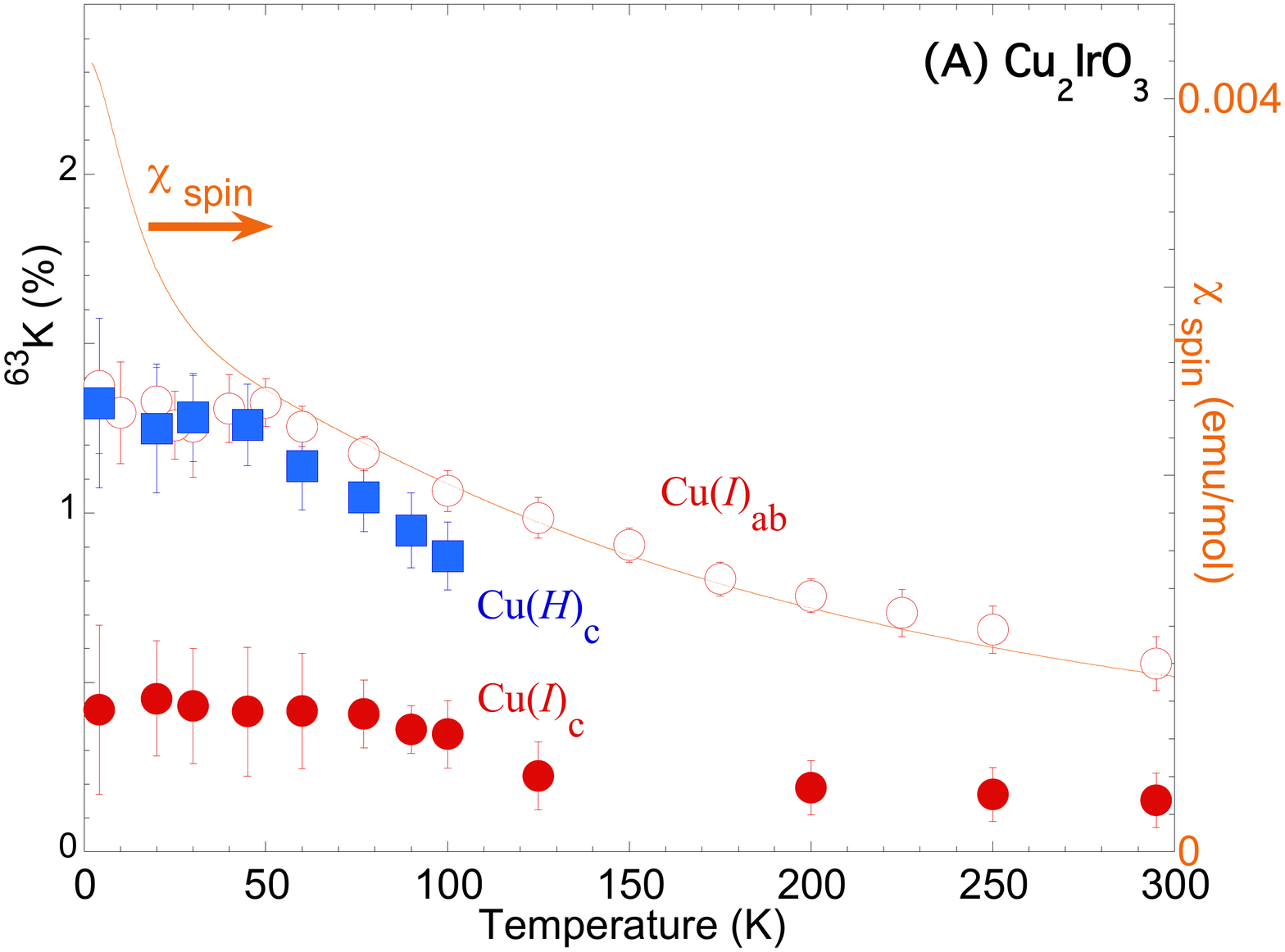}
		\includegraphics[width=3.2in]{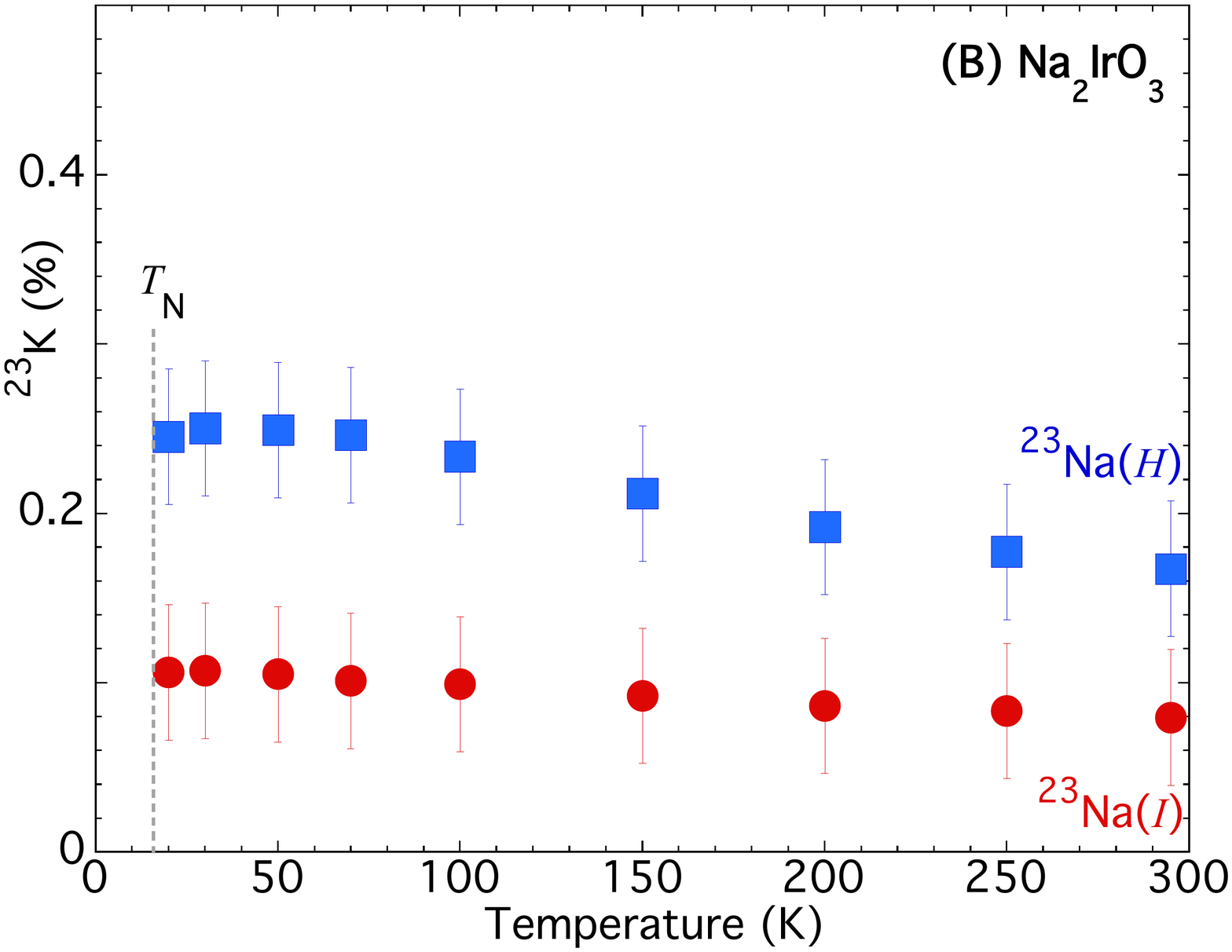}
		\caption{(A) The temperature dependence of the intrinsic local spin susceptibility deduced as $^{63}K$, in comparison to the bulk averaged susceptibility $\chi_{spin}$ measured in 5 T (right axis).  Notice that the large upturn observed below $\sim 30$~K for the bulk $\chi$ data is absent in $^{63}K$.  (B) The temperature dependence of the intrinsic local spin susceptibility deduced as $^{23}K$ in Na$_2$IrO$_3$, as determined from the NMR lineshapes measured for a powder sample in Fig.\ 17.}
		\label{fig:shift}
	\end{center}
\end{figure}

\begin{figure}
	\begin{center}
		\includegraphics[width=3.2in]{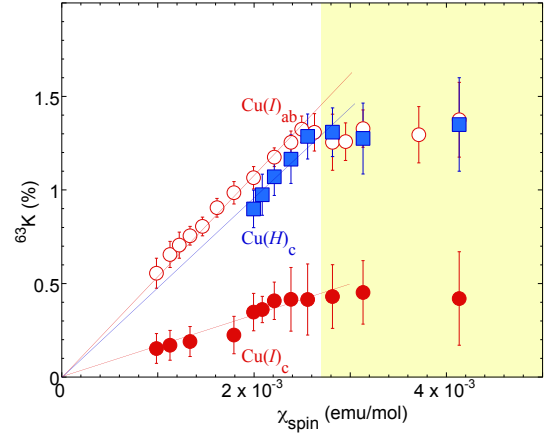}
		\caption{The plot of $^{63}K$ vs. the spin contribution $\chi_{spin}$ for Cu$_2$IrO$_3$, choosing temperature as the implicit parameter.  The solid lines represent the best liner fit through the origin.  $\chi_{spin}$ is enhanced by defect spin contributions in the yellow shaded region, where the linearity breaks down.}
		\label{fig:Kchi}
	\end{center}
\end{figure}

\section{Discussions}
The activation behavior of $1/T_{1}$ is generically observed when the spin excitation spectrum represented by $S({\bf q},E)$ has a gap in the low energy sector.  For example, the collective spin-singlet state realized in the two-leg spin ladders SrCu$_2$O$_3$ \cite{Azuma1994} and La$_6$Ca$_8$Cu$_{24}$O$_{41}$ \cite{Imai1998} has a spin excitation gap as large as $\Delta/k_{B} \sim$ 450 K ($\sim J_{H}/2$ along the rung formed by a pair of Cu ions), and exhibits an activation behavior of $1/T_{1}$ at both $^{63}$Cu \cite{Azuma1994,Imai1998} and $^{17}$O \cite{Imai1998} sites below $\sim$500 K.  Our $1/T_{1}$ results in Fig.\ 7 is therefore consistent with the presence of a gap $\Delta/k_{B} = 175 \pm 30$~K ($\Delta \sim 15$~meV) in the primary mode of the intrinsic spin excitation spectrum in Cu$_2$IrO$_3$.  Note that this gap is comparable with the Ising interaction energy scale estimated above, $| J_{K} | = 17 \sim 30$~ meV.  In the pure Kitaev model, earlier theoretical analyses showed that the dispersion of the spin 1/2 excitations of fractionalized Majorana fermions may indeed have a gap as large as $\Delta_{Majorana} \sim 1.2J_{K}$ \cite{Knolle2015, Yoshitake2017}.  It is therefore tempting to associate our finding with the gapped excitations expected for Majorana fermions.  

	Broad high-energy spin excitations centered around ${\bf q}$ = ${\bf 0}$ with a gap $\Delta \sim 1.2 | J_{K} |$ was previously reported for $\alpha$-RuCl$_3$ by inelastic neutron scattering, but the low energy sector is filled by additional spin-wave like excitations \cite{Banerjee2016,Banerjee2017Science}.  In the present case, our observation of the faster component in $1/T_{1}$ in the recovery curve $M(t)$ indicates the presence of additional, spatially inhomogeneous relaxation mechanism(s).  Theoretically, excitations of a pair of gauge fluxes with total spin 1 could short-circuit the slower NMR relaxation process in the pure Kitaev model, and such gauge flux excitations might morph into the spin-waves (also with net spin 1) under the perturbation caused by the Heisenberg exchange term.  It is not clear what roles the gauge fluxes and/or spin waves may be playing in our results.  On the other hand, the growing distribution of $P(1/T_{1})$ toward $T_f$ seems to suggest that the faster component of $1/T_{1}$ originates primarily from extrinsic defect spins.  
	 
	It is important to distinguish our observation of the gapped behavior in zero magnetic field from earlier observations of a small field-induced gap in $\alpha$-RuCl$_3$ \cite{Baek2017, Zheng2017,Jansa2018}.  To test the potential influence of the applied magnetic field on the gapped behavior observed for Cu$_2$IrO$_3$, we measured $1/T_{1}$ at the lower satellite transition of the $^{63}$Cu(I$_{1,2}$) sites in $B_{ext}=9$~T applied along the aligned ab-plane. In this field geometry, $1/T_{1}$ probes the fluctuations of $h_{perp}$ along both the ab-plane and c-axis.  We summarize the NMR results with $\times$ symbols in Figs.\ 6A and 7A.  Comparison with the NQR results indicates that the gapped behavior of $1/T_{1}$ does not depend significantly on the magnetic field, at least up to 9~T. 
	
	A potential caveat in our interpretation of the gapped behavior of $S({\bf q},E)$ is that, in principle, $1/T_{1}$ can be suppressed at low temperatures if the form factor satisfies $A_{hf}({\bf q}_{AF}) = 0$ at a staggered vector ${\bf q}_{AF}$, where $S({\bf q}_{AF},E_{n})$ grows due to short-range spin-spin correlations \cite{Millis1990}.  In this alternate scenario, Ir spins on the entire honeycomb planes continuously grow strong short-range order down to $T_f$ without developing long-range order, resulting in diminishing $1/T_{1}$ due to geometrical cancellation of $A_{hf}({\bf q}_{AF})$ by the high symmetry at Cu sites.  
	
To illustrate the underlying mechanism in this alternate scenario, we present the possible short-range order patterns of Ir spins \cite{Choi2012} in Fig. 1B-C.   Unlike Na$_2$IrO$_3$ with strong zig-zag short range order, $1/T_1$ in Cu$_2$IrO$_3$ does not grow below $\sim 150$~K.  Accordingly, we can rule out strong zigzag short range order for Cu$_2$IrO$_3$. However, if Ir spins develop extremely strong N\'eel-type short-range spin-spin correlations shown in Fig. 1C, the transferred hyperfine field $h_{perp}$ from 6 nearest-neighbor Ir sites of $^{63}$Cu(H) would cancel out, because the sign of $h_{perp}$ alternates between six nearest-neighbors.  Although the location of the $^{63}$Cu(I$_{1,2}$) sites is somewhat shifted from the mid-point between two adjacent Ir sites, $h_{perp}$ should also nearly cancel out for $^{63}$Cu(I$_{1,2}$) sites.  Thus, in principle, N\'eel-type short-range spin-spin correlations can suppress $1/T_1$ in Cu$_2$IrO$_3$.    However, we recall that such perfect cancellation of $h_{perp}$ expected for $1/T_{1} \sim 0$ observed here generally requires extremely long spin-spin correlation length $\xi$.  For example, in the case of $^{17}$O NMR measurements in the square-lattice Heisenberg antiferromagnet Sr$_2$CuO$_2$Cl$_2$ with no spin excitation gap above $T_N$ \cite{Thurber1997}, $\xi$ reaches tens of lattice spacings \cite{Greven1995} when $1/T_{1}$ is completely suppressed due to the high symmetry at the $^{17}$O sites.  In view of the frustrated geometry of Ir spins in the present case, validity of such a scenario starting from as high as $\sim150$~K (comparable to $\sim |J_{K}|/k_{B}$, and much higher than $|J_{H}|/k_{B}$) seems somewhat remote.  In addition, the intrinsic uniform spin susceptibility deduced from $^{63}K$ in Figure 11A does not exhibit any hint of a downturn below $\sim$150 K, which is expected for such strong quasi-two-dimensional short-range N\'eel order.  It is also worth recalling that $^{17}$O as well as $^{63}$Cu NMR $1/T_{1}$ measurements of two-leg spin-ladders in La$_6$Ca$_8$Cu$_{24}$O$_{41}$ successfully singled out the gapped spin excitation mode at $\sim$40 meV, even though $A_{hf}({\bf q})$ cancels at the $^{17}$O sites \cite{Imai1998}.   
	  
Next, we wish to address the nature of defect spins and their influence on the NMR properties.  As noted in section I, recent XANES measurements for Cu$_2$IrO$_3$ showed that up to $\sim$1/3 of $^{63}$Cu(H) sites are occupied by Cu$^{2+}$ ions rather than Cu$^+$ ions.  On the other hand, the linear fit of the $^{63}K$ vs. $\chi$ plot in Fig.\ 12 extrapolates to the origin for both $^{63}$Cu(H) and $^{63}$Cu(I$_{1,2}$) sites.  This means that the van Vleck contribution to the Knight shift is vanishingly small, as expected for Cu$^+$ ions with filled 3d orbitals.  If the observed $^{63}$Cu(H) NMR signals arise from Cu$^{2+}$ ions, the intercept with the vertical axis of the extrapolated linear fit would be as large as 0.3\% to 1.5\% due to the temperature independent van Vleck contribution for Cu$^{2+}$ ions \cite{Mila-Rice}.  
	
The lack of Cu$^{2+}$ contributions in our $^{63}$Cu(H) NMR data may be easily understood, if we recall that NMR signals at paramagnetic cation sites are generally not observable due to extremely fast NMR relaxation rates.  In fact, $^{63}$Cu NQR and NMR signals in the paramagnetic state of the aforementioned ZnCu$_3$(OH)$_6$Cl$_2$ \cite{Imai2008}, CuO \cite{Itoh1990}, SrCu$_2$O$_3$ \cite{Azuma1994}, La$_6$Ca$_8$Cu$_{24}$O$_{41}$ \cite{Imai1998}, and Sr$_2$CuO$_2$Cl$_2$ \cite{Thurber1997} etc. are observable only because exchange narrowing effects induced by strong antiferrmagnetic exchange interaction $J_{H}/k_{B}$ of the order of 200 to 1700~K between cation sites suppresses the NMR relaxation rates, which certainly would not be the case for the Cu$^{2+}$ defect spins in Cu$_2$IrO$_3$.  We therefore conclude that the primary contributions of Cu$^{2+}$ defect spins to our NMR results are through the line broadening effects for the observable $^{63}$Cu(H) and $^{63}$Cu(I$_{1,2}$) NMR signals below $\sim$50 K and the enhancement of $1/T_{1}$ in their vicinity.  The latter leads to a large distribution of $1/T_{1}$, as seen in $P(1/T_{1})$ in Figs.\ 7B and 8.  
	
\section{Summary and conclusions}	
We have reported comprehensive $^{63}$Cu NQR and NMR measurements in Cu$_2$IrO$_3$, a proximate Kitaev QSL material that does not undergo magnetic long-range order.  We showed that the intrinsic uniform spin susceptibility $\chi_{spin}$ of Cu$_2$IrO$_3$ shows nearly identical behavior as in the paramagnetic state of antiferromagnetic Na$_2$IrO$_3$ above $T_{N} \sim 17$~K.  On the other hand, the low energy sector of Ir spin excitations reflected on the temperature dependence of $1/T_1$ is very different in Cu$_2$IrO$_3$.  $1/T_1$ shows activation behavior and is qualitatively different from the conventional behavior exhibited by antiferromagnetic Na$_2$IrO$_3$.  Our finding strongly suggests that the low energy sector of the spin excitation spectrum in Cu$_2$IrO$_3$ is dominated by a mode with a gap comparable to the magnitude of Ising interaction $| J_{K} |$, which may be related to gapped excitations expected for fractionalized Majonara fermions perturbed by Heisenberg exchange term.  In such an idealized scenario of the Kitaev lattice, it is not clear what roles the excitations of a pair of gauge fluxes might play.  Analogous dilemma was previously encountered in the analysis of the spin excitations of $\alpha$-RuCl$_3$, because the low energy sector is dominated by magnons rather than gauge fluxes \cite{Banerjee2017Science}.  From the measurements of $1/T_1$ alone, we cannot entirely rule out an alternate scenario that Ir spin-spin correlation develops strong N\'eel type short range order with divergently long spin-spin correlation length.  However, that would normally require suppression of $\chi_{spin}$ in quasi two-dimensional Heisenberg antiferromagnets, and we did not find such a signature in our $^{63}K$ results.

\begin{acknowledgments}
The authors thank S.-S. Lee, R. Moessner, M.J. Graf, I. Kimchi, S. M. Winter, N. Perkins, Y. Motome, Y. B. Kim, and C. Varma  for helpful discussions and communications.  T.I. is supported by NSERC.  F.T. is supported by NSF under DMR-1708929.  P.M.S. is supported by The Rice University Consortium for Processes in Porous Media.
\end{acknowledgments}

\appendix
\section{\label{sec:level1}Fitting procedures for $1/T_{1}$}
The results of the nuclear magnetization $M(t)$ for the $1/T_1$ measurements at $^{63}$Cu(I$_1$), $^{63}$Cu(I$_2$) and $^{63}$Cu(H) sites are qualitatively similar, and we applied the same fitting procedures.  In what follows, we will focus on the results for the $^{63}$Cu(I$_1$) sites that have the highest signal to noise ratio.  In Figure 13A-B, we summarize $T_1$ recovery curves $M(t)$ observed for the $^{63}$Cu(I$_1$) site at selected temperatures.  For the $^{63}$Cu NQR measurements between the $I_{z} = \pm 3/2$ and $\pm 1/2$ states in zero magnetic field, magnetic transition would result in the exponential form of the recovery function \cite{Andrew1961, Narath1967}, 
\begin{equation}
M(t) = M_{o}-A \cdot exp(-3t/T_{1}),
\end{equation} 
where $1/T_{1}$, $M_o$ and $A$ are the fitting parameters.  We found that this single exponential form fits the recovery data nearly perfectly above $\sim$60 K, as shown by dashed straight lines. 
	
\begin{figure}[b]
	\begin{center}
		\includegraphics[width=3.2in]{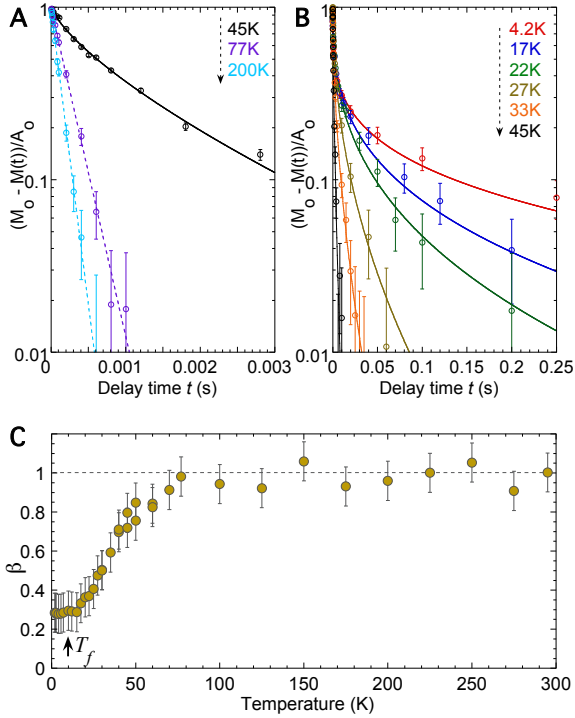}
		\caption{(A-B) The $^{63}$Cu NQR $T_1$ recovery curves observed at representative temperatures for the Cu(I$_1$) sites, presented in the form of $(M_{o} - M(t))/A$.  Dashed straight lines are the best exponential fit with eq.(A1), while the solid curves are the best phenomenological stretched exponential fit with eq.(A2).  (C) The stretched exponent $\beta$ obtained from the fit with eq.(A2) choosing $\beta$ as a free parameter. }
		\label{fig:T1stretch}
	\end{center}
\end{figure}	

\begin{figure}[b]
	\begin{center}
		\includegraphics[width=3.2in]{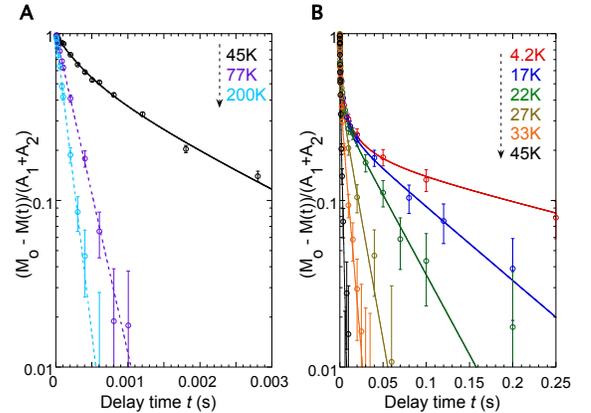}
		\caption{(A-B) Solid curves represent the two component fit with eq.(A3) above (left panel A) and below (right panel B) 45~K, presented in the form of $(M_{o}-M(t))/(A_{1}+A_{2})$.  Notice that after the initial quick decay caused by $1/T_{1,fast}$, there is a long, straight tail associated with the slow component $1/T_{1,slow}$.  Dashed straight lines through the 77~K and 200~K data points are the best exponential fit with eq.(A1).}
		\label{fig:T1double}
	\end{center}
\end{figure}

Below $\sim$60 K, $1/T_{1}$ begins to exhibit a distribution and hence the fit with eq.(A1) becomes progressively poorer.  As a remedy, one can introduce a phenomenological stretched exponent $\beta$ introduced for ranfomly diluted antiferromagnets \cite{Thayamballi1980, Itoh1986},
\begin{equation}
M(t) = M_{o}- A \cdot exp(-[3t/T_{1,str}]^{\beta}).
\end{equation} 
If there is no distribution in the relaxation process, the stretched fit would give $1/T_{1,str} = 1/T_{1}$ and $\beta = 1$.  As summarized in Fig.\ 13C, the stretched exponent begins to deviate from $\beta =1$ below $\sim 60$~K as the intrinsic relaxation process slows down and becomes susceptible to the faster defect contributions.  $1/T_{1,str}$ results are approximate, but tend to elucidate the sample averaged behavior, as proven by the $P(1/T_{1})$ results in Fig.\ 8 based on inverse Laplace transform of $M(t)$.  That is, the center of gravity of the distribution function $P(1/T_{1})$ agrees fairly well with $1/T_{1,str}$.

Our goal, however, is not merely to find the sample averaged behavior.  A better approach to capture the intrinsic slow component  $1/T_{1,slow}$ is the more traditional two-component fit of $M(t)$, 
\begin{eqnarray}
M(t) = M_{o}-A_{1} \cdot exp(-3t/T_{1,slow}) \nonumber \\
-A_{2} \cdot exp(-[3t/T_{1,fast}]^{\beta})
\end{eqnarray} 
Here, $M_o$, $A_1$, the intrinsic slow component $1/T_{1,slow}$, $A_{2}$, the defect-induced fast component $1/T_{1,fast}$ are the fitting parameters.  We also introduced the phenomenological stretched exponent $\beta$ in the second term to improve the fit (the estimated values of $\beta$ from eq.(A3) are similar to those from eq.(A2)).  It is worth noting that analogous two component fit was previously applied successfully to diluted antiferromagnets Mn$_{1-x}$Zn$_{x}$F$_2$ \cite{Itoh1986}.  (The site-dilution effect caused by non-magnetic Zn$^{2+}$ ions in Mn$_{1-x}$Zn$_{x}$F$_2$ results in distributed slow component of $1/T_1$ in the vicinity of Zn$^{2+}$ ions, while the fast intrinsic component obeys the single exponential form.)  

Aside from the precedent, a concrete justification of the two component fit in the present case comes from the fact that inverse Laplace transform of $M(t)$ indeed shows the presence of two peaks in a wide temperature range below $\sim$60 K down to $T_f$, as shown in Fig.\ 8.  Fig.\ 14A and B show the normalized recovery curves $(M_{o} - M(t))/(A_{1}+A_{2})$ plotted in a semi-logarithmic scale; the distinct slow component $1/T_{1,slow}$ manifests itself as a straight segment after the initial quick decay due to $1/T_{1,fast}$.   The $1/T_{1}$ results below $\sim$60 K plotted in Figs.\ 5B, 6A, 7A, and 7B with filled symbols are $1/T_{1,slow}$ based on the fit with eq. (A3).  The open symbols in Figs. 5B and 7A represent $1/T_{1,str}$ from the fit with eq. (A2) or $1/T_{1,fast}$ from the fit with eq. (A3).

 %================================
\section{\label{sec:level1}$^{63,65}$Cu NMR measurements in high magnetic field $B_{ext}$}
 \subsubsection{Field sweep $^{63,65}$Cu NMR lineshapes} 
	In Fig.\ 15, we compare the $^{63,65}$Cu NMR lineshapes measured at 90.307~MHz for uniaxially aligned and un-aligned powder samples at 10 K while sweeping the external magnetic field.  For the purpose of determining the NMR Knight shift accurately, we measured only the $I_{z} = -1/2$ to +1/2 central peak region in a fixed magnetic field of $B_{ext}=9$~T while sweeping the frequency, as shown in Fig.\ 9A-C. 
	
\begin{figure}
	\begin{center}
		\includegraphics[width=3.2in]{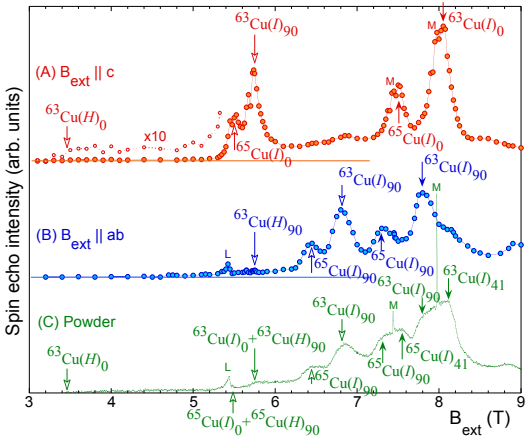}
		\caption{Field sweep $^{63,65}$Cu NMR lineshapes observed at 10~K at a fixed frequency of 90.307~MHz for (A-B) a uni-axially aligned powder and (C) an unaligned powder.  Key features associated with $^{63}$Cu ($^{65}$Cu) isotope are marked with downward (upward) arrows.  The subscripted numbers 0, 41, and 90 denote the angle $\theta$ between $B_{ext}$ and the crystal c-axis.  Filled and open arrows distinguish the features arising from the $I_{z} = +1/2$ to -1/2 central transition and $I_{z} =\pm 1/2$ to $\pm 3/2$ satellite transition.  Peaks designated with the letter M (or L) are background signals from $^{63}$Cu and $^{65}$Cu metal parts (or $^{7}$Li) in the sample environment.   }
		\label{fig:CuNMR}
	\end{center}
\end{figure}	
	
	The lineshape for the unaligned powder sample in Fig.\ 15C exhibits a typical powder pattern with a double horn structure arising from the $I_{z} = -1/2$ to +1/2 central transition of the $^{63}$Cu and $^{65}$Cu isotopes, as identified by filled green arrows.  The horn at the lower field side originates from nuclear spins in the grains whose primary principal axis of the electric field gradient (EFG) tensor (i.e. the crystal c-axis) forms a right angle $\theta \sim 90^\circ$ with the external magnetic field, whereas the horn at the higher field side has $\theta \sim 41 ^{\circ}$.  (Their relative locations are reversed in the frequency sweep lineshapes in Fig.\ 9.)  These structures are smeared by magnetic line broadening at 10~K caused by defect spins.    
	
	When we apply $B_{ext}$ along the ab-plane of the aligned powder sample, the $\theta \sim 90 ^{\circ}$ peak is enhanced, as seen in both Fig.\ 15B and Fig. 9C.  On the other hand, when we apply $B_{ext}$ along the aligned c-axis, a new peak emerges in the middle of the $\theta \sim 90 ^{\circ}$ and $\theta \sim 41 ^{\circ}$ peaks as seen in Fig.\ 15A and Fig. 9B; this is because the effects of the nuclear quadrupole interaction on the NMR frequency shift, $\Delta \nu_{Q}^{(2)}$, vanishes for both Cu(I) and Cu(H) along $\theta \sim 0 ^{\circ}$ due to their negligibly small asymmetry of the EFG tensor.  Since the NMR properties of the Cu(I$_1$) and Cu(I$_2$) sites are very similar, we were unable to clearly resolve their broad NMR lines.  
	
	At 10 K, the Cu(H) central peaks are hidden by stronger signals arising from the Cu(I$_1$) and Cu(I$_2$) sites, because the quadrupole effects are very strong at Cu(H) sites and the NMR line is broadened by defect spins.  Notice, however, that the $^{63,65}$Cu NMR signal in Fig.\ 15 extends below the lower bound expected for the $I_{z} = \pm 1/2$ to $\pm 3/2$ satellite transitions of Cu(I) sites, and diminishes instead at the edge near $B_{ext} \sim 3.4$ T as expected from $^{63}\nu_{Q} \sim 52$~MHz at $^{63}$Cu(H) sites.

 \subsubsection{Magnetic field effect on $1/T_{1}$} 
	We tested the potential effect of external magnetic field $B_{ext}$ on Ir spin dynamics by measuring $1/T_{1}$ for the aligned powder sample in $B_{ext}=9$~T, using the $\theta \sim 90 ^{\circ}$ $I_{z} = \pm 1/2$ to $\pm 3/2$ lower satellite transition for the $^{63}$Cu(I) sites.  The standard recovery function in this case takes the form of \cite{Andrew1961,Narath1967}
\begin{eqnarray}
M(t) = M_{o}-A[0.4~exp(-6t/T_{1}) \nonumber \\  
+0.5~exp(-3t/T_{1})+0.1~exp (-t/T_{1})].	
\end{eqnarray} 		
In analogy with the two component fitting function eq.(A3) for the central transition, we used the two component fit for the satellite transition as well,  
\begin{eqnarray}
M(t) = M_{o}-A_{1}[0.4~exp(-(6t/T_{1})) \nonumber \\  +0.5~exp(-(3t/T_{1})) 
 +0.1~exp (-(t/T{1}))] \nonumber \\
-A_{2}[0.4~exp(-(6t/T_{1,fast})^{\beta}) +0.5~exp(-(3t/T_{1,fast})^{\beta}) \nonumber \\
+0.1~exp (-(t/T_{1,fast})^{\beta})], \nonumber \\  
\end{eqnarray} 
as shown in Fig.\ 16.  The high field NMR $1/T_{1}$ results presented with $\times$ symbols in Fig. 6A and 7A, are nothing but $1/T_{1,slow}$ from the first term in eq.(B2), and the temperature dependence is nearly identical with the NQR results.  $1/T_{1,fast}$ and $\beta$ estimated from eq.(B2) are also similar to the NQR results.

\begin{figure}
	\begin{center}
		\includegraphics[width=3.2in]{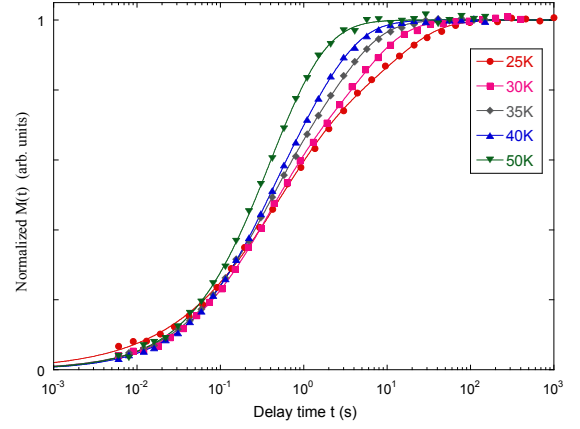}
		\caption{Normalized $T_1$ recovery curves $M(t)$ measured in $B_{ext}=9$~T at selected temperatures for the $I_{z} = \pm 1/2$ to $\pm 3/2$ lower satellite transition of $^{63}$Cu(I) sites.  Solid curves represent the two component fit with eq. (B2).  }
		\label{fig:CuNMRT1}
	\end{center}
\end{figure}

\section{\label{sec:level1}$^{23}$Na NMR measurements in Na$_2$IrO$_3$}

\begin{figure}
	\begin{center}
		\includegraphics[width=3.2in]{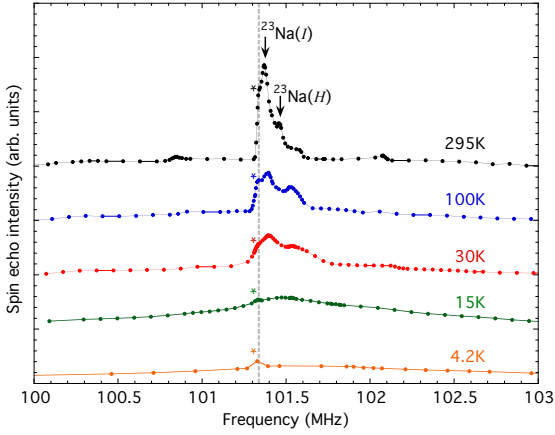}
		\caption{$^{23}$Na powder NMR lineshapes observed at $B_{ext} =9$~T for Na$_2$IrO$_3$.  The grey dashed vertical line marks the unshifted $^{23}$Na frequency $^{23}\nu_{o}$.  The hump marked with * is an unshifted $^{23}$Na signal from a small amount of magnetically inert impurity phase.}
		\label{fig:NaNMR}
	\end{center}
\end{figure}

\begin{figure}
	\begin{center}
		\includegraphics[width=3.2in]{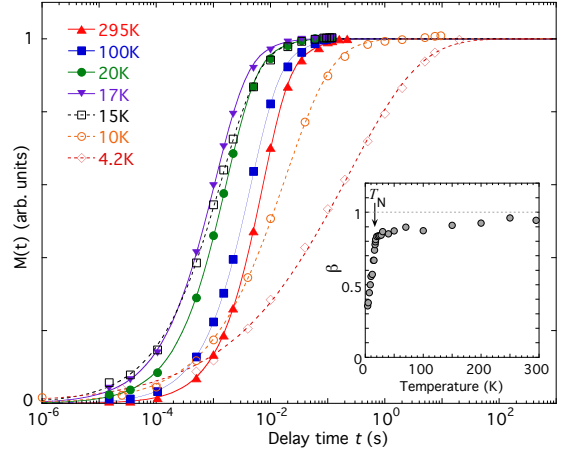}
		\caption{$T_1$ recovery curves for the nuclear spin $I_{z} = +1/2$ to -1/2 central transition of the inter-layer $^{23}$Na(I) sites of Na$_2$IrO$_3$ at select temperatures. Solid and dashed curves represent the fit with eq.(C1) with the stretched exponent $\beta$ above and below $T_N$, respectively.  The magnitude of $\beta$ is close to 1 above $T_N$, as shown in the inset.}
		\label{fig:NaT1fit}
	\end{center}
\end{figure}

In Fig.\ 17, we summarize $^{23}$Na (nuclear spin $I = 3/2$) powder NMR lineshapes measured in $B_{ext}=9$~T.  At 295 K, we observed two central peaks from nuclear spin $I_{z} = +1/2$ to -1/2 transition.  We assign the main peak centered at $\sim 101.37$~MHz to the interlayer $^{23}$Na(I) sites.  The smaller side peak at higher frequency $\sim 101.46$~MHz with larger Knight shift should be attributed to the less abundant sodium site within the honeycomb plane, referred to as $^{23}$Na(H).  We were also able to resolve shoulders from $I_{z} = \pm 3/2$ to $\pm 1/2$ satellite transitions at $\sim 102.1$~MHz and $\sim 100.8$~MHz.  An additional small peak/shoulder at $\sim 101.3$~MHz, marked with *, is at the non-shifted frequency, and should be attributed to a small amount of non-magnetic impurity phase.  The lineshape broadens dramatically below $T_{N}\sim 16.5$~K due to the emergence of static hyperfine magnetic fields from the ordered Ir moments.

We also measured $1/T_{1}$ at the relatively isolated $^{23}$Na(I) central peak.  We summarize the recovery curves $M(t)$ in Fig.\ 18, and the temperature dependence of $1/T_{1}^{Na}$ in Fig.\ 6B.  Despite the superposition of NMR signals arising from the satellite transitions of the $^{23}$Na(I) sites as well as $^{23}$Na(H) sites, the fit of the recovery curve $M(t)$ with the standard form for the central transition \cite{Andrew1961,Narath1967}, 
\begin{eqnarray}
M(t) = M_{o}-A \cdot [0.9~exp(-(6t/T{1})^{\beta}) \nonumber \\
+ 0.1~exp(-(t/T{1})^{\beta})],	
\end{eqnarray} 
was good with fixed $\beta = 1$ down to $\sim$30 K.  To improve the fit below $T_{N}$, we introduced the stretched exponent $\beta$ as summarized in the inset of Fig.\ 18.  Since the value of $\beta$ remains close to $\sim 1$ above $T_N$, the qualitatively important aspects of $1/T_{1}$ results in Fig.\ 6B do not depend significantly on the details of the fit, with or without $\beta$.

%\nocite{*}

%\bibliography{Takahashi_Cu2IrO3_190827}% Produces the bibliography via BibTeX.


\end{document}